\documentclass[epj,nopacs]{svjour}
\usepackage{graphicx}

\def\i{\mathrm{i}}

\def\d{\mathrm{d}}
\def\half{{\textstyle{1\over2}}}
\def\thalf{{\textstyle{3\over2}}}
\def\fhalf{{\textstyle{5\over2}}}
\def\h{{\scriptscriptstyle{1\over2}}}

\def\vec#1{\mbox{\boldmath$#1$}}
\def\CG#1#2#3#4#5#6{C^{#5#6}_{#1#2#3#4}}
\def\U{\uparrow}               
\def\D{\downarrow}

\begin{document}

\title{Eta and kaon production in a chiral quark model}

\author{%
B.~Golli \inst{1}\thanks{\email{bojan.golli@ijs.si}}
\and
S.~\v{S}irca \inst{2}\thanks{\email{simon.sirca@fmf.uni-lj.si}}
}

\institute{%
Faculty of Education,
              University of Ljubljana and J.~Stefan Institute,
              1000 Ljubljana, Slovenia 
\and
Faculty of Mathematics and Physics,
              University of Ljubljana and J.~Stefan Institute,
              1000 Ljubljana, Slovenia}

\date{\today}

\abstract{%
We apply a coupled-channel formalism incorporating quasi-bound 
quark-model states to calculate pion scattering into $\eta N$, 
$K\Lambda$ and $K\Sigma$ channels, as well $\eta p$, $\eta n$, 
$K^+\Lambda$, and $K^0\Sigma^+$ photo-production processes.
The meson-baryon and photon-baryon vertices are determined in 
a SU(3) version of the Cloudy Bag Model.
Our model predicts sizable amplitudes in the $P_{11}$, $P_{13}$, 
$P_{33}$ and $S_{11}$ partial waves in agreement with the latest
MAID isobar model and the recent partial-wave analyses of 
the Bonn-Gatchina group.  We are able to give a quark-model 
explanation for the apparent resonance near 1685~MeV in the 
$\eta n$ channel.}

\maketitle

\section{Introduction}

The analysis of inelastic scattering and photoproduction of 
mesons in the framework of quark models which puts strong 
limits on the meson-baryon and photon-baryon vertices
provides important insight into the mechanism of resonance 
formation and represents a useful guidance to choosing relevant 
degrees of freedom and parameters in more phenomenological 
methods used in the partial wave-analyses of large data sets.
The channels involved in production of $\eta$ and $K$ mesons 
give the opportunity to assess the validity of the SU(3) symmetry 
of the quark model in describing the decay of the resonances.
Furthermore, these channels provide invaluable information
on the resonances that are weakly coupled to the $\pi N$
channel or overwhelmed by large background and therefore
hardly observed in pion production.

We use a coupled-channel formalism incorporating 
qua\-si-bound quark-model states to calculate the pion- and 
photo-production amplitudes in a unified approach.
The meson-baryon and photon-baryon vertices are determined 
in a chiral quark model.
We use a SU(3) extended  version of the Cloudy Bag Model (CBM)
which includes also the  $\rho$ and $\omega$ mesons.
The method has been described in detail in our previous
papers~\cite{EPJ2005,EPJ2008,EPJ2009,EPJ2011,EPJ2013}
where we have analysed the scattering and 
electro-production amplitudes in different partial waves.
We studied the $\eta N$ channel in the $S_{11}$ partial
wave~\cite{EPJ2011}, pointing out its important 
contribution in the region of the $N(1535)$ resonance; we also 
considered the $K\Lambda$ channel  in this partial wave.
In the present work we systematically analyse the partial
waves with sizable contributions to the $\eta N$, $K\Lambda$ 
and $K\Sigma$ decay channels: in addition to $S_{11}$, these 
are the $P_{11}$, $P_{13}$ and $P_{33}$ partial waves.

In this work we have decided not to adjust the parameters
of the model to fit particular resonances but keep them at the 
values used in the ground-state calculation; only the positions
of the $K$-matrix poles  and the mixing angles between 
the quark model states with the same parity, isospin and 
angular momentum are considered as adjustable parameters.
As we discuss in sec.~\ref{EM}, the calculation of different
amplitudes proceeds in the same way as in the Cloudy Bag Model.
Based on the vast experience gathered in this model we can 
estimate the reliability of our calculation to (20--30)~\%.
An agreement on this level with the corresponding partial-wave 
analysis (PWA) in any given resonance region may be deemed 
reasonable.
However, since no adequate analytic tool
has been used to quantitatively verify these estimates,
they should be accepted with caution, in particular at higher
energies.

Experimentally, photoproduction of $\eta$ mesons has been 
studied extensively at all modern real-photon facilities.  
Angular distributions, total cross-sections and several 
polarization observables have been measured in broad energy 
ranges; see, for example, 
Refs.~\cite{crede05,bartalini07,crede09,williams09,mcnicoll10,werth14} 
for select results on the $\eta p$ channel and 
Refs.~\cite{werth14,rebre05,jaegle08,jaegle11} 
for the $\eta n$ channel.
The photoproduction of the $\eta$ meson has attracted a particular 
interest after the GRAAL Collaboration~\cite{Kuznetsov07}
observed a narrow structure at $W\approx1685$~MeV in the 
$\gamma n\to\eta n$ reaction which was, however, absent 
in the $\eta p$ channel.
Azimov {\em et al.}~\cite{Azimov05} were the first to discuss the 
possibility that the structure could belong to a partner 
of the $\Theta^+$ pentaquark in the exotic antidecouplet of baryons.
More conventional explanations have attributed the peak to the 
threshold effect of the $K\Sigma$ channel~\cite{Doering10},
interference of the nearby $S_{11}$, $P_{11}$ and $P_{13}$
resonances~\cite{Shyam08}, constructive and destructive 
interference of the two lowest $S_{11}$ resonances in the $\eta n$ 
and $\eta p$ channels, respectively, as anticipated in the 
framework of the Giessen model~\cite{Shklyar07,Shklyar13} 
as well as in the Bonn-Gatchina analysis~\cite{BoGa09,BoGa15}.
In the framework of the constituent-quark model coupled to the 
pseudoscalar meson octet the (non)appearance of the peak was 
related to different EM multipoles responsible for excitation 
in either of the two channels~\cite{Zhong11}.
A similar model applied to $\eta$ production~\cite{Saghai09,Saghai10}
has revealed the important role of the lowest $S_{11}$,
$P_{13}$, $D_{13}$ and the lowest two $F_{15}$ resonances.
The photoproduction of $\eta$ mesons off the proton has been
extensively studied in a gauge-invariant chiral unitary approach
\cite{MaiPLB11,MaiPRD12} with dynamically generated $S11$ 
resonances, resulting in a good agreement with data for the $E_{0+}$ 
pion and  $\eta$ amplitudes.

Photo-production of the $K^+\Lambda$ final state on the proton
has also been a major focus of several recent experimental efforts,
resulting in a large body of data on partial and total cross-sections,
as well as polarization observables
\cite{bock94,tran98,glan04,brad04,nabb04,sumi06,ller07,ller09,crack10}.
Theoretically, pion- and photon-induced $K\Lambda$ production has been 
studied in different approaches in order to establish the 
main mechanism governing its behaviour at low energies close 
to the kaon production threshold.
Li {\em et al.}~\cite{Li95,Li96} studied the photoproduction of
kaons in a chiral constituent-quark model and stressed
the importance of the $S$-wave resonances.
Steininger and Mei{\ss}ner~\cite{Ulf97} and Borasoy 
{\em et al.}~\cite{Ulf06} studied the threshold behaviour of the kaon 
photoproduction in the framework of chiral perturbation theory. 
The Giessen group found sizable contributions from the 
$S_{11}$(1650), $P_{13}$(1720) and $P_{13}$(1900) 
resonances~\cite{Shklyar05} and later upon also from the 
$S_{11}$(1535), with the background comparable to the resonant 
terms~\cite{Shyam10}.
The Gent group~\cite{Gent2012} also concluded that the $S_{11}(1650)$ 
and $P_{13}(1720)$ resonances contribute most to photo- and 
electro-production in this channel.
Mart~\cite{Mart10,Mart11} considered only the $S_{11}$(1650) 
resonance and found that the dominant contribution arises 
from $t$- and $u$-channel background.
The Kent group has included the $\eta N$ and $K\Lambda$ channels 
in their recent multichannel partial-wave 
analysis~\cite{Manley12b,Manley12a}.

Due to similar energy ranges and particle identification requirements,
experimental investigations of the photoproduction of the $K\Sigma$
final state on the proton have gone hand-in-hand with the $K\Lambda$
studies, and we now have a rich data set on both partial and total
cross-sections: see, for example, 
Refs.~\cite{bock94,tran98,glan04,brad04,nabb04} for the $K^+\Sigma^0$
channel and \cite{goers99,lawall05,klein05,cast08,ewald12,aguar13} 
for the $K^0\Sigma^+$ channel.
Considering the low-energy photoproduction of the $K\Sigma$ 
channels close to the threshold, the Giessen group~\cite{Shklyar13} 
have pointed out the interference of the two low-lying $S_{11}$ 
resonances, while the $P_{11}$ and $P_{13}$ waves turn out
to become important at somewhat higher energies; 
for the isospin-$3/2$ channels the $P_{33}$ wave seems 
to be considerably stronger than the $S_{31}$ wave,
in contrast to the previous result~\cite{Shyam10}. 
The relatively small cross-section for the $K^0\Sigma^+$ 
compared to the $K^+\Sigma^0$ channel is explained by the 
interference of the resonant and background terms.
The Bonn-Gatchina group~\cite{BoGa13ks} found similar results
regarding the importance of the $S_{11}$ and $P$-wave resonances.
Mart~\cite{Mart14}, on the other hand, considered the 
$P_{11}$(1710), $P_{13}$(1720), $D_{13}$(1700) and $D_{33}$(1700) 
resonances.

A unified description of elastic and inelastic channels 
including also the production of $\eta$ mesons and kaons has
been studied in the dynamical coupled-channel models of the  
Bonn-J\"ulich group~\cite{Roenchen13} and the Osaka-Tokyo-Argonne 
group~\cite{Kamano13} in order to systematically extract several 
parameters of the resonances below 2~GeV. 
The  Bonn-J\"ulich results for the pion-induced reactions
confirm the dominant role of the $S_{11}$ wave at lower energies 
in the isospin-$1/2$ channels, and the $P_{33}$ wave in the 
isospin-$3/2$ ones.

In the next section we briefly overview the calculation of the pion- 
and photo-production amplitudes in our coupled-channel approach for 
the $K$ matrix and introduce the SU(3) version of the CBM.
We discuss the resonances included in the  $P_{13}$ partial 
wave which has not been considered in our previous work.
We also discuss in more detail the $P_{11}$ wave resonances 
since we have now included several new channels with respect 
to our previous calculation.

In sect.~3 we present the scattering amplitudes involving
$\eta N$, $K\Lambda$ and $K\Sigma$ channels, and in sect.~4 
the photoproduction amplitudes.
We discuss in detail the difference between the $\eta$ 
photoproduction amplitudes on the proton and on the neutron.
We give a possible quark-model mechanism for this difference
which may explain the apparent resonant structure in the total 
cross-section near 1685~MeV.  
The results are summarized in sect.~5.

\section{\label{model} The model}

\subsection{\label{Kformalism} The coupled channel formalism}

In chiral quark models in which mesons couple linearly to the quark 
core  the elements of the $K$~matrix in the basis with good total 
angular momentum $J$ and isospin $I$  take the form \cite{EPJ2008}:
\begin{equation}
 K_{M'B'\,MB}^{JI} =  -\pi\mathcal{N}_{M'B'}
   \langle\Psi^{MB}_{JI}||V^{M'}(k)||\Psi_{B'}\rangle\,,
\label{defK}
\end{equation}
where $V^{M'}(k)$ stands for the quark-meson vertex of the
underlying quark model, $\Psi_{B'}$ is the
baryon state in the $M'B'$ channel, and
$\mathcal{N}_{MB} = \sqrt{\omega_{M} E_{B} / (k_{M} W)}$.
Here $W$ is the invariant energy, $E_B$ is the baryon energy
and $\omega_{M}$ and $k_{M}$ are the meson energy and momentum.
The expression (\ref{defK}) differs from the expression for the 
corresponding $T$ matrix in that the incoming state is replaced 
by the principal-value state $\Psi^{MB}_{JI}$; in the following, 
all integrals are assumed in the princi\-pal value sense.
The quark-model quasi-bound states $\Phi_{\mathcal{R}}$ are included 
through the following ansatz:
\begin{eqnarray}
|\Psi^{MB}_{JI}\rangle &=& \mathcal{N}_{MB}\left\{
    [a^\dagger(k_M)|\Psi_B\rangle]^{JI} 
+
 \sum_{\mathcal{R}}c_{\mathcal{R}}^{MB}|\Phi_{\mathcal{R}}\rangle
\right. \nonumber\\ && \left. 
+ \sum_{M'B'}
   \int {\d k\>
       \chi^{M'B'\,MB}(k,k_M)\over\omega_k+E_{B'}(k)-W}\,
      [a^\dagger(k)|\Psi_{B'}\rangle]^{JI}\right\}\,,
\nonumber\\
\label{PsiH}
\end{eqnarray}
where the first term represents the free meson ($\pi$, $\eta$, 
$K$, $\dots$) and the baryon ($N$, $\Delta$, $\Lambda, \ldots$) 
and defines the channel, the next term is the sum over 
{\em bare\/} three-quark states, while the third term 
describes meson clouds around different isobars.
The principal-value states (\ref{PsiH}) are normalized as
\begin{equation}
   \langle\Psi^\alpha(W) |
                     \Psi^\beta(W')\rangle
  = \delta(W-W') \left[\delta_{\alpha,\beta} 
  + {\mathbf{K}^2}_{\alpha,\beta}\right]\,.
\label{normPV}
\end{equation}
They are not orthonormal; the orthonormalized states
are constructed by inverting the norm.

The meson amplitudes $\chi^{M'B'\,MB}(k,k_M)$ and the coefficients 
$c_{\mathcal{R}}^{MB}$ are obtained from the Kohn variational 
principle leading to a set of coupled equations: equations of 
the Lippmann-Schwinger type for the meson amplitudes
\begin{eqnarray}
&&   \chi^{M'B'\,MB}(k,k_M) 
   = -\sum_{\mathcal{R}}{c}^{MB}_{\mathcal{R}}\, {V}^{M'}_{B'\mathcal{R}}(k)
\nonumber\\ && 
       +\>\> \mathcal{K}^{M'B'\,MB}(k,k_M)
\nonumber\\ && 
+ \sum_{M''B''}\int\d k'\,
  {\mathcal{K}^{M'B'\,M''B''}(k,k')\chi^{M''B''\,MB}(k',k_M)
  \over \omega_k' + E_{B''}(k')-W}\,,
\nonumber\\
\label{eq4chi}
\end{eqnarray}
and equations for the coefficients $c_{\mathcal{R}}^{MB}$:
\begin{eqnarray}
 (W-M_{\mathcal{R}}^{(0)}) {c}^{MB}_{\mathcal{R}}
   &=& V^M_{B\mathcal{R}}(k_M) 
\nonumber\\
&+& \sum_{M'B'}\int\d k\,
    {\chi^{M'B'\,MB}(k,k_M) V^{M'}_{B'\mathcal{R}}(k)\over 
     \omega_k + E_{B'}(k)-W}\,.
\nonumber\\
\label{eq4c}
\end{eqnarray}
Here
\begin{equation}
  \mathcal{K}^{M'B'\,MB}(k,k')
=  
  \sum_{B''} f_{BB'}^{B''}\,
  {\mathcal{V}_{B''B'}^{M'}(k') \, 
   \mathcal{V}_{B''B}^{M}(k)
   \over \omega_k+\omega_k'+E_{B''}(\bar{k})-W}\,,
\label{kernel}
\end{equation}
\begin{eqnarray*}
  f_{AB}^C  &=& \sqrt{(2J_A+1)(2J_B+1)(2I_A+1)(2I_B+1)}\,
\\
&\times&   W(l_BJ_AJ_Bl_A;J_C,J)W(i_BI_AI_Bi_A;I_C,I)\,,
\end{eqnarray*}
$l_A$, $i_A$, $\ldots$ are the meson, $J_A$, $I_A$, $\ldots$ 
are the baryon angular momenta and isospins,
${V}^{M}_{B\mathcal{R}}(k)$ are the bare matrix elements of the 
quark-meson interaction between the baryon state $B$ and the
bare three-quark state $\Phi_{\mathcal{R}}$, and $M_{\mathcal{R}}^{(0)}$ 
is the energy of the bare state.
The corresponding dressed vertices obey the equation
\begin{equation}
 \mathcal{V}^M_{B\mathcal{R}}
   = {V}^{M}_{B\mathcal{R}}(k)
+ \sum_{M'B'}\int\d k'\,
  {\mathcal{K}^{MB\,M'B'}(k,k')\,\mathcal{V}^{M'}_{B'\mathcal{R}}(k')
  \over \omega_k' + E_{B'}(k')-W}\,.
\label{eq4calV}
\end{equation}

Solving the coupled set of equations can be considerably
simplified by using a separable approximation for the kernel
of the form
\begin{eqnarray}
&&  \kern-42pt {1 \over \omega_k+\omega_k'+E_{B''}-W} \approx    
\nonumber \\ &&
   {(\omega_M + \omega_{M'} + E_{B''} - W)
\over 
   (\omega_k+E_{B''}-E_{B'})(\omega_k'+E_{B''}-E_B)}
\label{separable}
\end{eqnarray}
where $W = E_B + \omega_M = E_{B'} + \omega_{M'}\,$.
The approximation on the RHS of (\ref{separable}) coincides with the 
exact expression on the LHS when either of the two pions is on-shell.
Furthermore, the dressed vertices in (\ref{kernel}) can be replaced 
by the bare ones in those cases in which the mesons are only weakly 
coupled to baryons.
It turns out that in the case of the P33 and P11 partial wave,
in which the $p$-wave pions strongly couple to the quark core,
solving the integral equations alters the coupling constants by 
some 50~\% with respect to the bare quark values; in the case of 
$s$- and $d$-wave pions as well as other mesons, the modification
typically remains at the level of 5~\% to 10~\%. 
Let us note that neglecting the integrals in (\ref{eq4chi}), 
(\ref{eq4c}) and (\ref{eq4calV}) corresponds to the so-called Born 
approximation for the $K$ matrix.

The meson amplitudes are proportional to the half off-shell matrix 
elements of the $K$-matrix,
\begin{equation}
   K_{M'B'\,MB}(k,k_M)  = \pi\,\mathcal{N}_{M'B'}\mathcal{N}_{MB}\,
             \chi^{M'B'\,MB}(k,k_M) \,.
\label{chi2K}
\end{equation}
The procedure outlined above yields the final expression
\begin{eqnarray}
   K_{M'B'\,MB}(k,k_M) 
     &=& -\sum_{\mathcal{R}}
     {\mathcal{V}^M_{B\mathcal{R}}(k_M)
      \mathcal{V}^{M'}_{B'\mathcal{R}}(k)
      \over Z_{\mathcal{R}}(W) (W-W_{\mathcal{R}})}
\nonumber\\      && 
  + K^{\mathrm{bkg}}_{M'B'\,MB}(k,k_M)\,,
\label{sol4K}
\end{eqnarray}
where the first term represents the contribution of various 
resonances, while the second term originates in the non-resonant 
background processes; $Z_{\mathcal{R}}$ is the resonance wave function
normalization.

The scattering $T$ matrix is obtained by solving the
Heitler equation $T=K+\i KT$.

\subsection{\label{CBM} The underlying quark model}

The vertices are calculated in a version of the Cloudy Bag Model
extended to the pseudo-scalar, $\phi_a$, and vector, $\vec{A}_a$, 
SU(3) meson octet \cite{Thomas85}:
\begin{eqnarray}
H_{\mathrm{int}}
  &=& -\int\d\vec{r}\left[{\mathrm{i}\over 2f} \, \overline{q}\lambda_a 
    (\gamma_5 \phi_a + \vec{\gamma}\cdot\vec{A}_a)q\;\delta_S\right.
\nonumber\\
   &&  \left.  +{1\over4 f^2}\, \overline{q}\lambda_a \gamma^\mu q
        (\phi\times\partial_\mu \phi)_a\theta_V \right]\,, 
\nonumber \\
 && a=1,2,\ldots,8\;.
\label{Hint}
\end{eqnarray}
The model provides a consistent parametrization of the 
meson-baryon and photon-baryon coupling constants and form 
factors in terms of $f$ (equivalent to $f_\pi$) and the bag 
radius $R_\mathrm{bag}$.
We use $R_\mathrm{bag} = 0.83$~fm and $f = 76$~MeV, consistent 
with the values used in the ground-state calculations.
In addition, the bare masses of the resonances are also free 
parameters.
The last term in (\ref{Hint}) has not been included in our 
previous calculations.
It contributes to the background and influences most strongly 
the real part of the elastic $\pi N$ amplitude in the $P_{13}$ 
and $P_{31}$ partial waves.
Using the same value for $f$ as in the first term would result
in a too strong contribution of this term.
We therefore multiply this term by a factor $\half$.
Furthermore, the strength of $\rho$ meson coupling
should be reduced by a factor of 3, consistent with the value 
of the $\rho$ meson decay constant $f_\rho\approx200$~MeV.

\subsection{\label{EM} The photo- and electroproduction amplitudes}

In the case of photo- and electroproduction the strong vertex in 
(\ref{defK}) is replaced by the electromagnetic interaction vertex.
In chiral quarks models with the underlying Lagrangian, such as 
the Cloudy Bag Model used in our calculation, the electromagnetic 
currents are derived using the principle of minimal coupling.
The quark, pion and kaon contributions to the spatial part 
of the electromagnetic current read
\begin{eqnarray}
    \vec{j}_{EM}^{\,\,q} &=& \sum_{i=1}^3 \bar{\psi}\vec{\gamma}(i)\psi 
\left({1\over6} +  {\tau_0(i)\over2}\right)\,,
\label{jq}
\\
\vec{j}_{EM}^{\,\,\pi} 
 &=& \i\sum_t t\pi_t(\vec{r})\,\vec{\nabla}\pi_{-t}(\vec{r})\,,
\label{jpi}
\\
\vec{j}_{EM}^{\,\,K} 
 &=& \i\left[K^-(\vec{r})\,\vec{\nabla} K^+(\vec{r})
        -  K^+(\vec{r})\,\vec{\nabla} K^-(\vec{r})\right]\,.
\label{jK}
\end{eqnarray}
The resonant part of the electroproduction amplitude
can be cast in the form
\begin{equation}
{\mathcal{M}_{MB\gamma N}^\mathrm{res}}  =
\sqrt{\omega_\gamma E_N^\gamma \over \omega_\pi E_N }\,
{\xi\over\pi{\cal V}_{N\mathcal{R}}^\pi}\,
  \langle\widehat{\Psi}_{\mathcal{R}}|{V}_\gamma
                |\Psi_N\rangle\, {T_{MB\,\pi N}} \>,
\label{Vgamma2M}
\end{equation}
where $V_\gamma$ describes the interaction of the photon with the 
electromagnetic current and $\xi$ is the spin-isospin factor
depending on the considered multipole and the spin and isospin
of the outgoing hadrons.  
The resonance state $\widehat{\Psi}_{\mathcal{R}}$ is extracted from 
the components in the second and the third term in (\ref{PsiH}) 
that are proportional to the resonance pole $(W-W_{\mathcal{R}})^{-1}$;
it involves the bare-quark core and the meson cloud:
\begin{equation}
|\widehat{\Psi}_{\mathcal{R}}\rangle
=   
Z_{\mathcal R}^{-{1\over2}}\left[|{\Phi}_{\mathcal{R}}\rangle
   - \sum_{MB}\int{\d k\quad{\cal{V}}^M_{B{\mathcal R}}(k)
             \over\omega_k+E_B-W}\,
      [a^\dagger(k)|\Psi_B\rangle]^{JI}
\right].
\label{PsiR}
\end{equation}
Let us note that the calculation of the electromagnetic vertices
in our approach proceeds in the same way as in the
standard calculation in the framework of the  Cloudy Bag Model.

Although the corresponding  quark-level Lagrangian of the
Cloudy Bag Model respects electromagnetic gauge invariance, 
this may not be the case on the nucleon level.
Miller and Thomas \cite{gaugeCBM} have shown that the model 
respects gauge invariance in the Breit frame.
In this frame, no extra terms are needed to calculate
the nucleon electromagnetic form factors.
The fact that their result applies only to a specific frame 
should be a caveat to the reader, since this invariance should hold in
all frames. 
In the calculation of form-factors no attempt is usually made to
construct states with good linear momentum.
In Ref.~\cite{CBM4Delta} the amplitudes for electroproduction of 
the $\Delta$ resonance have been calculated using the Peierls-Thouless
method to eliminate the spurious center of mass motion.
This has altered the results obtained in the static approximation
by about 5~\% to 8~\% and enabled the authors to estimate
the error of neglecting recoil effects 
to $q^2/4m_N^2$, $q^2$ being the pion momentum squared.
Accordingly, we expect these deficiencies to map to our computed
amplitudes at a comparable level of uncertainty, not exceeding
20~\% below $q^2\approx 1~{\rm GeV}^2$.
A considerably larger disagreement might indicate a more severe
inconsistency in the quark-model picture of a specific resonance.
Another source of discrepancy could be a strong contribution of 
the background processes, in particular at higher energies around 
$\sim 1700$~MeV and above, not included or strongly underestimated 
in our calculation.

\subsection{Resonances in the $P_{13}$ partial wave}

In the quark model, the $N(1720)\thalf^+$ and 
$N(1900)\thalf^+$ resonances are obtained by exciting one $s$ quark 
to the $d$ orbit with either $j=\thalf$ or $j=\fhalf$.
(Note that the $p$-wave pseudoscalar mesons couple only to 
$j=\thalf$.)

The two physical resonances are linear combinations
of the spin doublet  and spin quadruplet states:
\begin{eqnarray}
|N(1720)\rangle &=& 
   \cos\vartheta_p|{\bf 70}, {}^2{\bf 8}, J=\thalf\rangle
-  \sin\vartheta_p|{\bf 70}, {}^4{\bf 8}, J=\thalf\rangle   \>, 
\nonumber\\
|N(1900)\rangle &=& 
   \sin\vartheta_p|{\bf 70}, {}^2{\bf 8}, J=\thalf\rangle
+  \cos\vartheta_p|{\bf 70}, {}^4{\bf 8}, J=\thalf\rangle   \>.
\nonumber\\
\end{eqnarray}
Here ${\bf 70}$ stands for the SU(6) flavour-spin multiplet 
and ${\bf 8}$ corresponds to isospin $I=\half$.
The detailed structure in terms of the $jj$-coupled states 
is given in Appendix~\ref{P13}.

\begin{figure}[h!]
\begin{center}
\includegraphics[width=65mm]{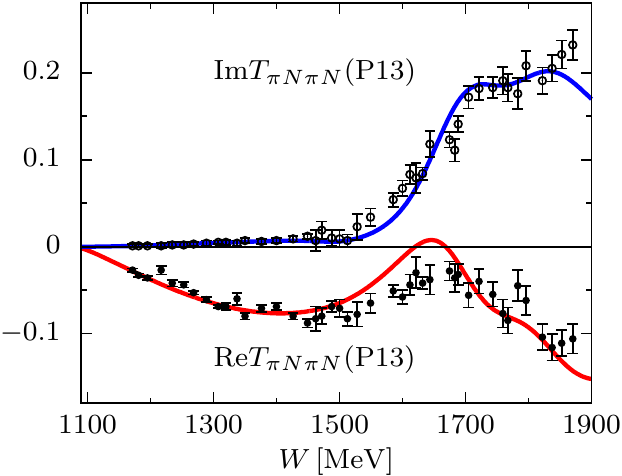}
\end{center}
\caption{The real and the imaginary part of the scattering
$T$ matrix for the $P_{13}$ partial wave.
The data points are from the SAID $\pi N\to\pi N$ partial-wave 
analysis~\cite{Arndt06,SAID}.}
\label{fig:TP13}
\end{figure}

The mixing originates from the gluon-quark and the meson-quark 
interaction.
The mixing due to gluon interaction has been investigated in 
the constituent-quark model 
\cite{Isgur77} and in the bag model~\cite{deGrand76b,Myhrer84b}.
In our model, the mixing arises also from meson-baryon interaction
as described in~\cite{EPJ2008}. 
The gluon interaction is not included in our model and the
corresponding mixing angle is taken as a free parameter.
The meson-loop contribution to the mixing angle is $W$-dependent.
In this partial wave the angle starts from $-15^\circ$, reaches 
$-30^\circ$  in the region of the lower resonance, and returns 
to $-15^\circ$ in the region of the upper resonance.
The $T$ matrix for the elastic channel is shown in fig.~\ref{fig:TP13}
in a coupled-channel calculation including $\pi N$, $\pi\Delta$,
$\sigma_{l=2}N$, $\eta N$, $\pi N(1440)$, $\rho_{l=0}N$, $K\Lambda$,   
$\rho_{l=2}N$, $K\Sigma$,  $\omega_{l=0}N$ and $\omega_{l=2}N$ channels.

\subsection{The Roper resonances revisited}

The $N(1440)\half^+$ resonance was extensively studied in our 
previous work~\cite{EPJ2008,EPJ2009,Pedro}, assuming 
the $\pi\Delta$ and $\sigma N$ inelastic channels.
In order to reproduce the behaviour of the scattering amplitudes 
up to $\approx 1800$~MeV, we included also the $N(1710)\half^+$
resonance assuming it decayed only into the $\sigma N$ channel.
If we want to study the production of strange mesons in the 
$P_{11}$ partial wave, we have to specify its quark configuration,
which is not obvious;
it may consist of the $(1s)^3$, $(1s)^2(2s)^1$, $(1s)^1(2s)^2$,
$(1s)^1(1p)^2$, $(1s)^1(1d)^2\ldots$ configurations or even a 
component with the excitation of the $\sigma$ cloud~\cite{Pedro}.
If we limit ourselves to the type of excitation in which
only a single $1s$ quark is raised to a higher orbit,
it suffices to specify only the admixture of $(1s)^2(2s)^1$
(and eventually of $(1s)^2(3s)^1$) configuration in the
resonance state.
We therefore assume the following structure of the two Roper
resonances:
\begin{eqnarray}
N(1440) &=& \cos\vartheta_R(1s)^2(2s)^1 - \sin\vartheta_R(1s)^1(1x)^2 \>,
\nonumber\\
N(1710) &=& \sin\vartheta_R(1s)^2(2s)^1 + \cos\vartheta_R(1s)^1(1x)^2 \>,
\label{Roper}
\end{eqnarray}
where $x$ stands for $l>0$ quark orbits not involved in the
transition matrix elements.

The $N(1440)\half^+$ and the nucleon also mix, which leads to the 
important nucleon-pole contribution to various amplitudes.
This mixing is calculated through the quark interaction with the
meson cloud (meson loops)  as discussed in~\cite{EPJ2008} and
does not introduce additional adjustable parameters.

The $\Delta(1232)\thalf^+$ and the $\Delta(1600)\thalf^+$ 
resonances are linear superposition of the $(1s)^3$ and 
$(1s)^2(2s)^1$ spin-3/2 configurations with the mixing 
angle calculated through the pion loops.

\subsection{Resonances in the $S_{11}$ partial wave}
The negative-parity resonances $N(1535)\half^-$ and 
$N(1650)\half^-$ have been extensively studied in our 
previous work~\cite{EPJ2011}.
Here the meson loops are less important compared to the $P$-wave
resonances and the mixing arises primarily through the
gluons~\cite{Isgur77,deGrand76b,Myhrer84b}:
\begin{eqnarray}
|N(1535)\rangle &=& 
   \cos\vartheta_s|{\bf 70}, {}^2{\bf 8}, J=\half\rangle
-  \sin\vartheta_s|{\bf 70}, {}^4{\bf 8}, J=\half\rangle \>,
\nonumber\\
|N(1650)\rangle &=& 
   \sin\vartheta_s|{\bf 70}, {}^2{\bf 8}, J=\half\rangle
+  \cos\vartheta_s|{\bf 70}, {}^4{\bf 8}, J=\half\rangle \>.
\nonumber\\
\label{S11LS}
\end{eqnarray}
The structure of these resonances in terms of the $jj$-coupled 
states involving quark excitation to $p_{1/2}$ and $p_{3/2}$ orbits
is given in~\cite{EPJ2011}.
In our previous work we made some adjustment of the quark-pion 
coupling strength in order to better reproduce the imaginary 
part of the amplitude in the elastic channel and in 
the $d$-wave $\pi\Delta$ channel.
In the present approach we keep the constants at their quark-model
values.
The mixing angle that best reproduces the $\eta N$ to $\pi N$
branching ratio is close to $-30^\circ$ and is only weakly
affected by the meson loops.

The isospin-3/2 partner 
$\Delta(1650)=|{\bf 70}, {}^2{\bf 10}, J=\half\rangle$
contributes very weakly to the amplitudes considered in
this work and will not be included in further analysis.
The same holds for the contribution of the $D$-wave resonances
treated in our previous work~\cite{EPJ2013}.


\section{Scattering amplitudes involving $\eta N$, $K\Lambda$ 
and $K\Sigma$ channels}

\begin{figure*}
$$\includegraphics[width=160mm]{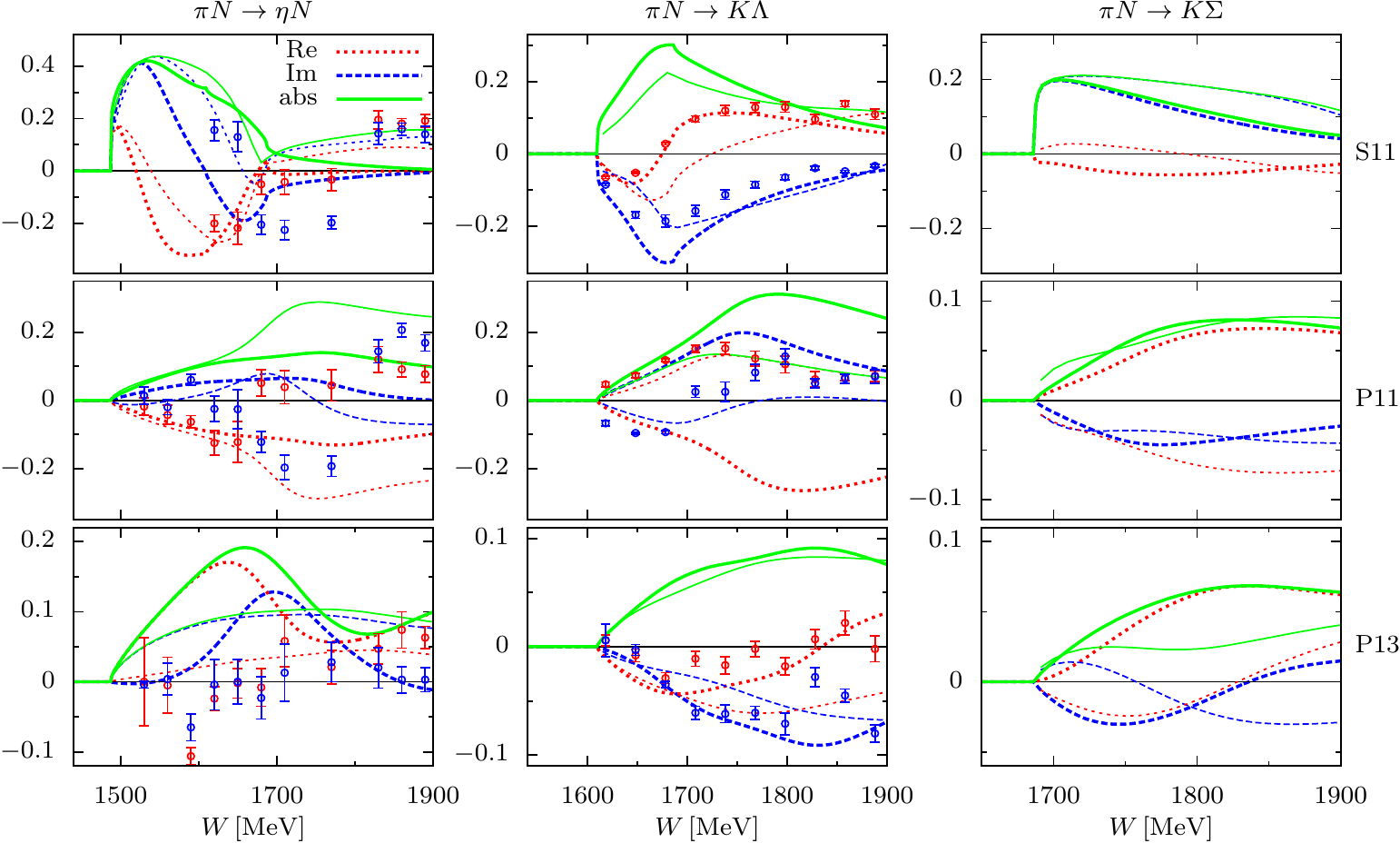}$$
\caption{(Color online.) 
The real and imaginary parts and the absolute values of the scattering 
$T$ matrix for the $\eta N$, $K \Lambda$  and $K\Sigma$ channels in the 
$S_{11}$, $P_{11}$ and $P_{13}$ partial waves (from top to bottom row).
The corresponding thin curves denote the 2014-2 solution of the 
Bonn-Gatchina group~\cite{BoGa-data}.  
The data points are from the single-energy analyses of 
Ref.~\cite{Manley92}.}
\label{fig:piN2strange}     
\end{figure*}

Figure~\ref{fig:piN2strange} shows the scattering amplitudes
for the $\pi N\to \eta N$, $\pi N\to K\Lambda$ and $\pi N\to K\Sigma$
reactions in the $S_{11}$,  $P_{11}$ and $P_{13}$ partial waves. 

For the $\pi N\to \eta N$ channel the dominant $S_{11}$ wave is very 
well reproduced compared to the Bonn-Gatchina analysis close to 
the threshold and in the region of the lower resonance; 
in the region of the upper resonance, the amplitudes follow the 
general trend and drop almost to zero.
In the $P_{13}$ partial wave our model predicts a too strong $\eta N$
amplitude, almost insensitive to the mixing angle. 
If we use a different value of the coupling constant,
consistent with the decay constant $f_\eta\approx1.2~f_\pi$,
the amplitude would drop by 20~\% but would still be too 
large compared to values from different partial-wave analyses.
The situation in the $P_{11}$ partial wave is rather intriguing
because of many distinct processes contributing to the amplitude:
the nucleon pole, the two Roper resonances and the $u$-channel process.
Our results in fig.~\ref{fig:piN2strange} are nonetheless not 
inconsistent with various partial-wave analyses.

The agreement of the calculated $\pi N\to K\Lambda$ amplitudes
with the partial-wave analyses is good and would further improve 
if we modified the quark-kaon coupling constant according to $f_K=1.2~f_\pi$.
While the $\eta N$ amplitude overshoots the PWA data in 
the $P_{13}$ partial wave, in the $K\Lambda$ case this happens 
in the $P_{11}$ wave.

The strength of the amplitudes in the $K\Sigma$ channel 
compares well with still rather uncertain analysis of the 
Bonn-Gatchina group, even without adjustment of the coupling 
constant (see fig.~\ref{fig:piN2strange}, right column).

\begin{figure}[h!]
\begin{center}
\includegraphics[width=70mm]{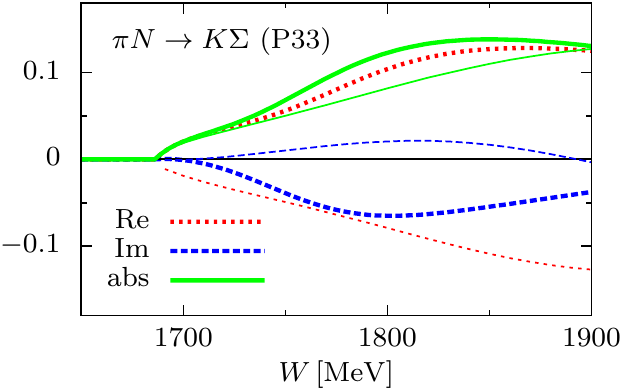}
\end{center}
\caption{(Color online.) The real and imaginary part of the 
scattering $T$ matrix for the $K\Sigma$ channel in the $P_{33}$ 
partial wave.  Notation as in fig.~\ref{fig:piN2strange}.}
\label{fig:P33}
\end{figure}

In fig.~\ref{fig:P33} the results for the $P_{33}$ partial 
wave are shown.
Also in this partial wave the magnitude of the scattering
amplitudes  agrees well with the Bonn-Gatchina analysis.

\section{Photoproduction}

Photoproduction involves the calculation of EM vertices 
in the framework of the same underlying quark model.
In general the photoexcitation amplitudes calculated in the 
CBM do not compare so favourably with experiment 
as in the case of strong vertices; already the magnetic 
moments of the nucleon are considerably underestimated if 
we insist on our value of the bag radius.
This should be kept in mind when assessing the quality of
our results.

In our previous work we calculated the electric multipoles 
from the charge density using the current conservation,
since -- in particular for the $E_{1+}$ multipole for the 
$\Delta(1232)$ excitation -- the results were less sensitive 
to the small variations of the model parameters \cite{PLB96}.
In the present work we strictly calculate both, the magnetic
as well as the electric multipoles, from the currents
(\ref{jq}) -- (\ref{jK}).

\subsection{\label{photoeta} $\eta p$ and $\eta n$ photoproduction}

Photoproduction of $\eta$ on the nucleon is particularly
interesting because of the presence of an apparent narrow 
resonance very close to the $K\Sigma$ threshold.
The intriguing feature of this structure is that it appears
only in the $\gamma n\to \eta n$ channel and is completely 
absent in the  $\gamma p\to \eta p$ channel.

\begin{figure*}[h!]
$$\includegraphics[width=160mm]{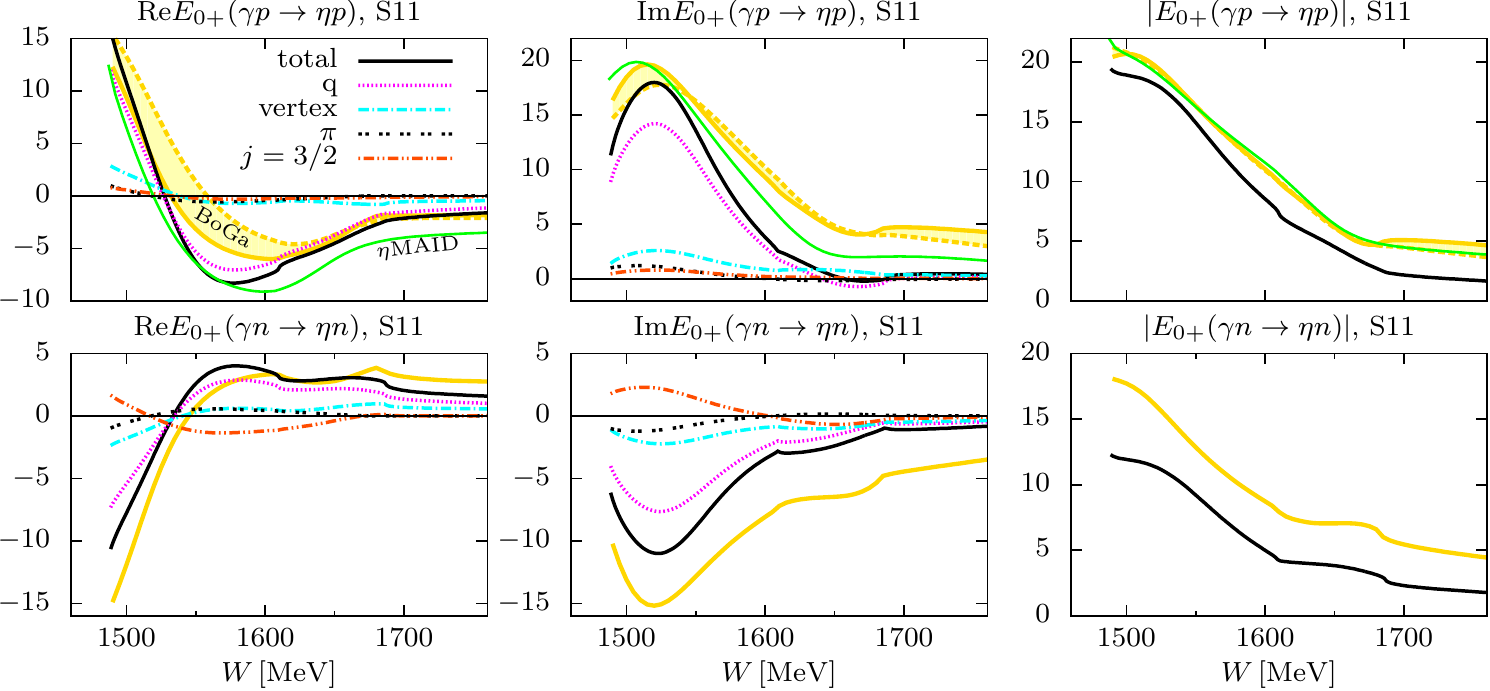}$$
\caption{(Color online.) 
The real and imaginary parts and the absolute values of the 
$E_{0+}$ multipole in $\eta$ photoproduction (in units of mfm). 
The Bonn-Gatchina solution (yellow) is multiplied  by $-1$; 
for $p\gamma$ the dashed lines correspond to the 2014-1 and 
the solid lines to the 2014-2 solution.  
For better visibility, the two Bonn-Gatchina solutions
(unless only one of them is available) are connected by shading.
The $\eta$MAID results (green) for $\gamma p$ are from \cite{ltpc}.
The quark, vertex-correction and pion-cloud contributions are shown
and, separately, the contribution from the $p_{3/2}$ orbit.}
\label{fig:photoeta1}
\end{figure*}

The photoproduction amplitudes are dominated by the $E_{0+}$
multipole and are reasonably well reproduced in our model
(see fig.~\ref{fig:photoeta1}); the neutron amplitude turns 
out to be somewhat weaker in the region of the lower 
resonance, while the imaginary parts of both amplitudes are 
underestimated in the region of the upper resonance.\footnote{%
In our previous work (see~\cite{EPJ2011}) there was an error
in the calculation of the neutron amplitudes, namely the
contribution involving the excitation of the $p_{3/2}$
orbit was taken with the wrong sign.}
The sign of our amplitudes agrees with those of the MAID 
analysis but is opposite to that of the Bonn-Gatchina group
 -- to facilitate the comparison we have reversed the sign
of the latter in fig.~\ref{fig:photoeta1}.

\begin{figure*}
$$\includegraphics[width=160mm]{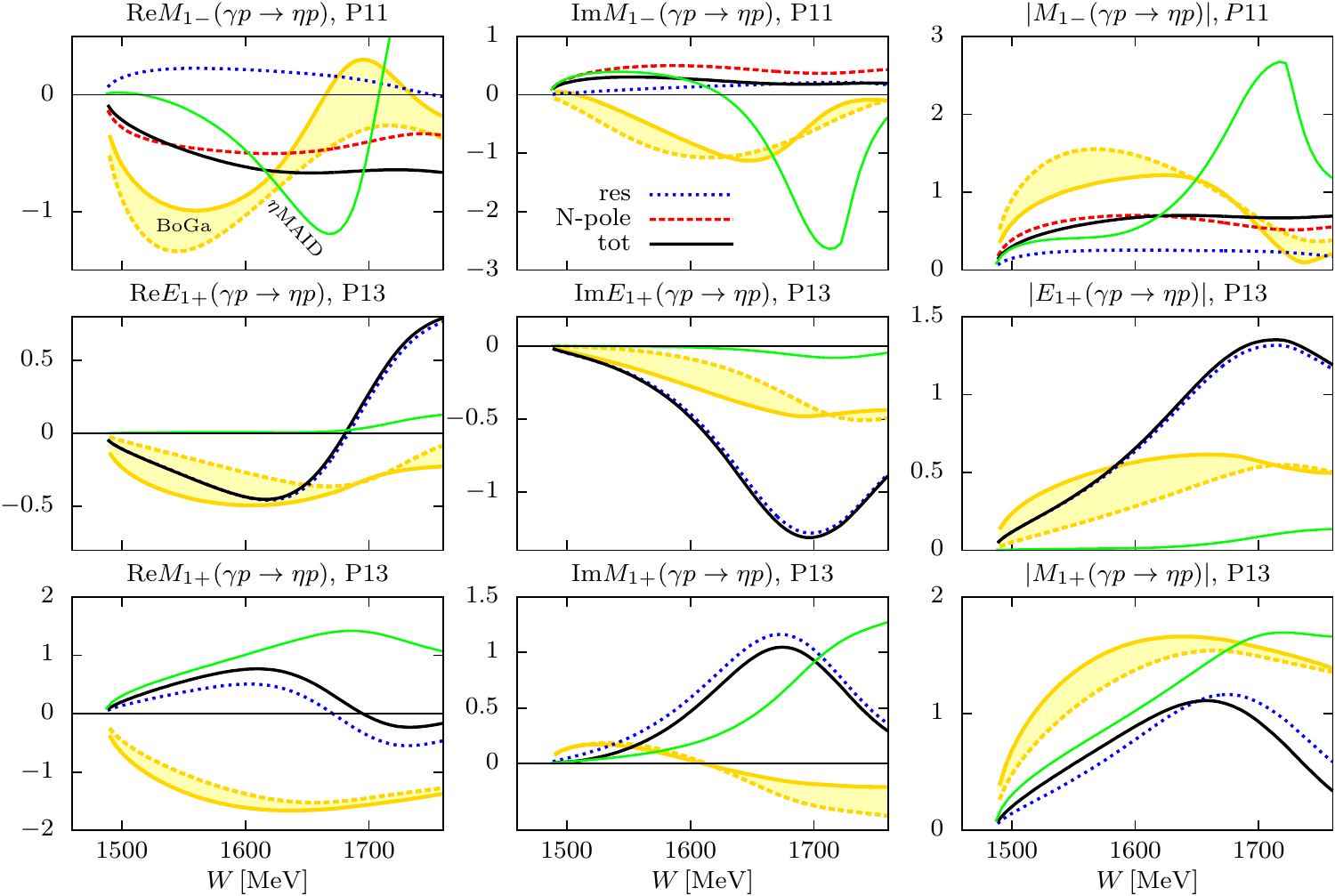}$$
\caption{(Color online.) The real and imaginary parts and 
the absolute values of the $M_{1-}$, $E_{1+}$ and $M_{1+}$ multipoles 
in $\eta$ photoproduction.  Notation as in fig.~\ref{fig:photoeta1}; 
the resonance (dots) and the nucleon-pole (dashes, in the $P_{11}$ wave) 
contributions are also shown.} 
\label{fig:photoeta2}
\end{figure*}

The $M_{1-}$, $M_{1+}$, $E_{1+}$ are less precisely determined
in the partial-wave analyses of the MAID and Bonn-Gatchina 
groups; our results in fig.~\ref{fig:photoeta2} reproduce 
the correct order of magnitude and overestimate the $E_{1+}$ 
amplitude, in accordance with a too large strong amplitude 
in this partial wave.

\begin{figure}[h!]
\begin{center}
\includegraphics[width=80mm]{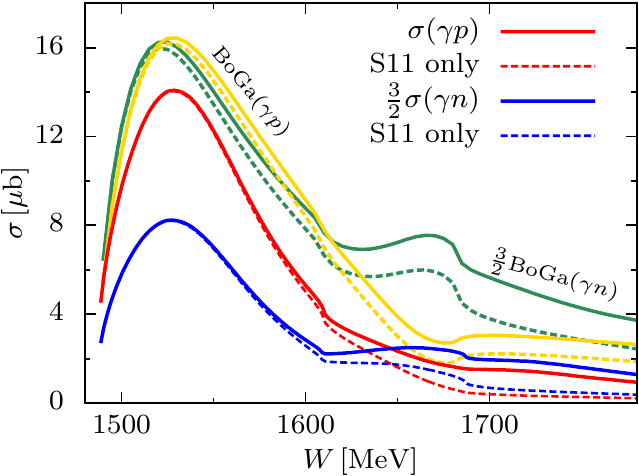}
\end{center}
\caption{(Color online.) The total cross-sections for 
$\gamma p\to \eta p$ and $\gamma n\to \eta n$ (multiplied 
by the conventional factor of ${3\over2}$).
Dashed lines: contribution of the $S_{11}$ partial wave.
The BoGa curves have been reconstructed from the Bonn-Gatchina 2014-2 
data set~\cite{BoGa-data}.}
\label{fig:sigmaeta}
\end{figure}

The total cross-section is displayed in fig.~\ref{fig:sigmaeta}
and compared to the results obtained from the Bonn-Gatchina 2014-2
analysis (which fits with great precision the experimental
cross-section).
The dominant $E_{0+}$ contribution is shown separately.
Our results do show a structure in the region of the $K\Sigma$
threshold in the $\gamma n$ channel, which is\break

\clearpage

\noindent 
absent in the $\gamma p$ 
channel, though its strength is lower compared to the experiment.%
\footnote{Readjusting the mixing angle could considerably improve
the quality of the $\gamma n$ result but would seriously deteriorate
the agreement for the other observables.}
Nonetheless, our analysis offers a possible and straightforward 
explanation of this structure in terms of the quark model: 
a combination of a peculiar property of the (relativistic) wave 
functions of the $S_{11}$ resonances and the presence of the 
$K\Sigma$ threshold.
Let us start from the states in (\ref{S11LS}) expressed in terms 
of the quark $p_j$ orbits~\cite{Myhrer84b}:
\begin{eqnarray}
|{}^4{\bf 8}_\h\rangle  &=& \phantom{-}{1\over3} \,|(1s)^2(1p_{3/2})^1\rangle 
                     + {\sqrt8\over 3} \,|(1s)^2(1p_{1/2})^1 \rangle \>,
\nonumber\\
|{}^2{\bf 8}_\h\rangle  &=& -{2\over3} \,|(1s)^2(1p_{3/2})^1\rangle 
                     + {\sqrt2\over 6}\,|(1s)^2(1p_{1/2})^1\rangle 
\nonumber\\
  &&                 + {\sqrt2\over 2}\,|(1s)^2(1p_{1/2})^1\rangle' \>,
\label{S11jj}
\end{eqnarray}
where the last two components with $p_{1/2}$ correspond to 
coupling the two $s$-quarks to spin 1 and 0, respectively;
the flavour (isospin) part is not written explicitly. 
In the strong part of the production amplitude, the $s$-wave 
$\eta$ meson couples only to the $p_{1/2}$ quark,
while the electric dipole ($E1$) photon couples to the 
$p_{1/2}$ component as well as to the $p_{3/2}$ component.
For the proton, however, the isoscalar part of the charge 
operator exactly cancels the isovector part in the case of
the first two components of both states in (\ref{S11jj}).
This is a general property and follows from the fact that the
flavour part in these two components corresponds to the
mixed symmetric state $\phi_{\mathrm{M,S}}$.\footnote{%
The property that the ${}^4{\bf 8}$ does not couple to the proton 
is known as the Moorhouse selection rule~\cite{Moorhouse1966}.}
The proton 
\break 

\noindent
therefore receives no contribution from the $1s\to 1p_{3/2}$ 
transition. 
This is not the case with the neutron which receives
contributions
from all components in (\ref{S11jj}).
The quark in the $1p_{3/2}$ orbit has a distinctly different radial
behaviour of the bispinor from that in the  $1p_{1/2}$ orbit 
(see Appendix~\ref{EMmulti}), which is reflected in a different 
momentum- and $W$-beha\-viour of the amplitudes.

\begin{figure}
\begin{center}
\includegraphics[width=75mm]{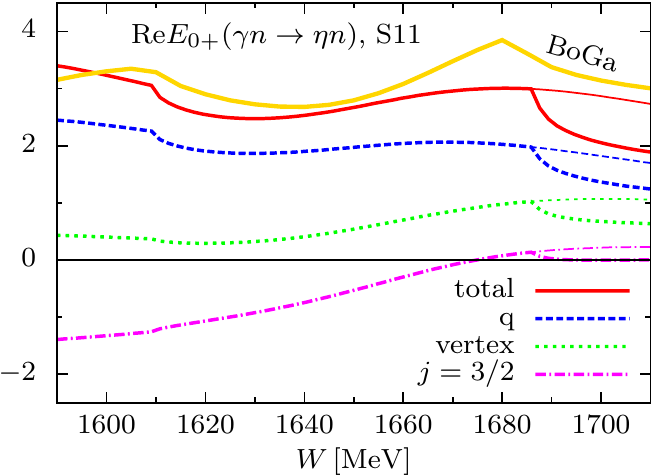}
\end{center}
\caption{(Color online.) Various contributions to the real part of 
the $E_{0+}$ amplitude: quark core (dashed), vertex correction (dotted),
contribution from the $p_{3/2}$ orbit (dash-dotted) in units of mfm. 
The Bonn-Gatchina results (thick yellow line at the top) is taken 
from the 2014-2 dataset and  multiplied by $-1$.
The corresponding thin curves denote the solution without the
$K\Sigma$ channel.}
\label{fig:S11enRe}
\end{figure}

To illustrate this point we display in fig.~\ref{fig:S11enRe}
various contributions to the real part of the $E_{0+}$ 
photoproduction amplitude in the $\eta n$ channel in the region 
between the $K\Lambda$ and the $K\Sigma$ threshold. 
The contribution of the $p_{3/2}$ orbit is negative and is rising
in this region, and influences the pure quark contribution and the
contribution from the vertex correction in such a way that the 
total amplitude becomes flat or even slightly increases.
When it reaches the $K\Sigma$ threshold it sharply drops as a 
consequence of the new channel opening (see fig.~\ref{fig:piN2strange}).
Had there not been a new channel, a wide and shallow structure 
would have been observed.
(We have chosen the case with the fixed mixing angle $-30^\circ$
since it sharpens the effect of the threshold compared to
a smoother behaviour in fig.~\ref{fig:sigmaeta}).

\subsection{$K^+\Lambda$ photoproduction}

\begin{figure*}
\begin{center}
\includegraphics[width=155mm]{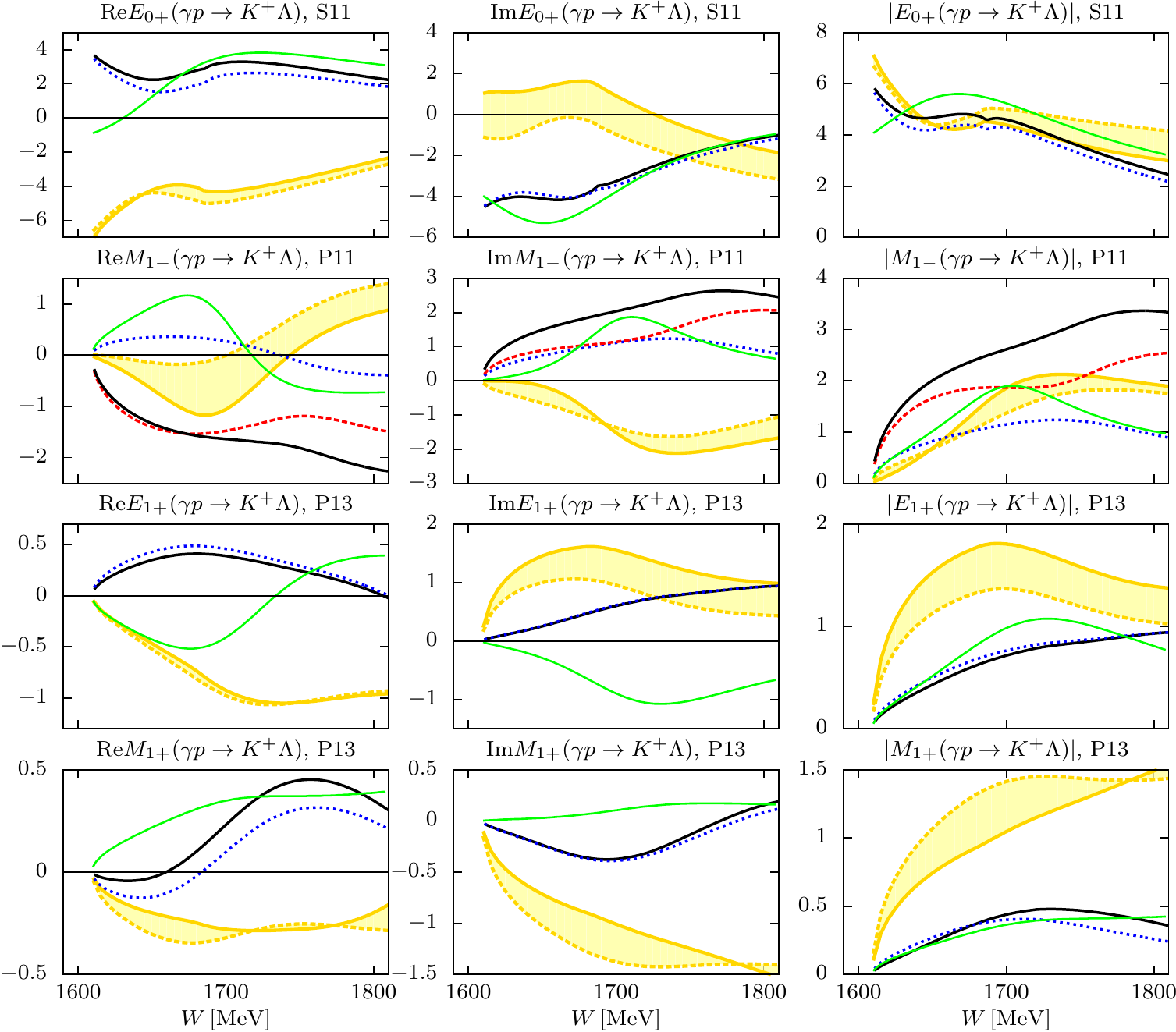}
\end{center}
\caption{(Color online.) The real and imaginary parts and the absolute 
values of multipoles dominating in  $K^+\Lambda$ photoproduction.
Notation as in fig.~\ref{fig:photoeta2}.}
\label{fig:photoKL}
\end{figure*}

\begin{figure}[h!]
\begin{center}
\includegraphics[width=70mm]{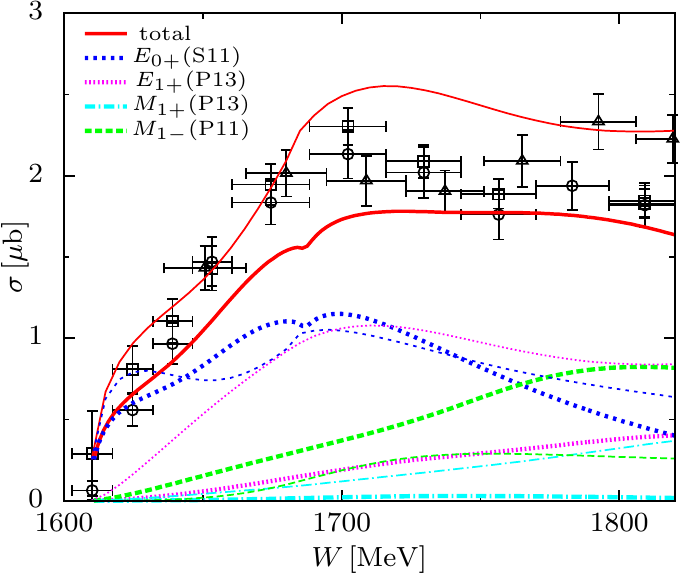}
\end{center}
\caption{(Color online.) Total cross-section for 
$\gamma p\to K^+\Lambda$.
The experimental data are from~\cite{bock94,tran98,glan04}.
The thin lines correspond to the cross-sections calculated from the
Bonn-Gatchina 2014-1 dataset.}
\label{fig:sigmaKL}
\end{figure}

At energies close to the threshold the $E_{0+}$ multipole 
represents the dominant contribution to $K$ photoproduction 
though not to the same extent as in the $\eta$ case 
(see fig.~\ref{fig:photoKL}).
Here our sign is again consistent with the Kaon MAID 
analysis~\cite{kaonMAID} but opposite to that of 
the Bonn-Gatchina group.  The strength of this multipole is, however, 
closer to the prediction of the Bonn-Gatchina group.
For the $P_{13}$ wave our results agree with MAID, while it seems that 
we obtain a too strong contribution in the case of the $P_{11}$ wave.
This contribution is sensitive to the admixture of the 
nucleon-pole term which remains important also at\break

\noindent
higher energies.
The trend can be seen in fig.~\ref{fig:sigmaKL} 
showing
different
contributions to the total cross-section.
The Giessen analysis of this channel~\cite{Shklyar05}
also predicts the dominant contribution from the $S_{11}$, 
$P_{13}$ and $P_{11}$ waves with the $P_{13}$ being stronger 
than the $P_{11}$ one, opposite to our calculation but 
in agreement with the Bonn-Gatchina prediction.
The total cross-section exhibits a surprisingly good
agreement with the experimental points which, admittedly,
could be a consequence of an overestimation of the
strong part (see fig.~\ref{fig:piN2strange})
and an underestimation of the EM amplitudes.

In our model the main contribution to the background originates 
from the $t$-channel; at low energies the coupled-channel effect 
involving the $\gamma p\to \pi N$ process and the $T_{\pi N\,K\Lambda}$ 
matrix element turns out to be dominant but still weaker than the 
resonant one.
We have not included the background involving the
$K^*$ and $K_1$ mesons.

\clearpage


\onecolumn

\subsection{$K^0\Sigma$ photoproduction}

We have decided to concentrate only on $K^0$ photoproduction
since the background processes in the case of $K^+$ 
photoproduction may considerably hinder the more interesting 
resonant contribution.

The amplitudes in this channel involve isospin-$1/2$ and
isospin-$3/2$ contributions~\cite{Mart14}:
$$
A(\gamma + p \rightarrow K^0\Sigma^+) = 
\sqrt{2}\left[A_p^{(1/2)} - {1\over3}\,A^{(3/2)}\right] \>,
\qquad
A(\gamma + n \rightarrow K^0\Sigma^0) = 
-A_n^{(1/2)} + {2\over3}\,A^{(3/2)}\,.
$$
In the case of the $E_{0+}$ multipole, the contribution from 
the $S_{31}$ partial wave turns out to be negligible in our model.

\begin{figure*}[h!]
\begin{center}
\includegraphics[width=155mm]{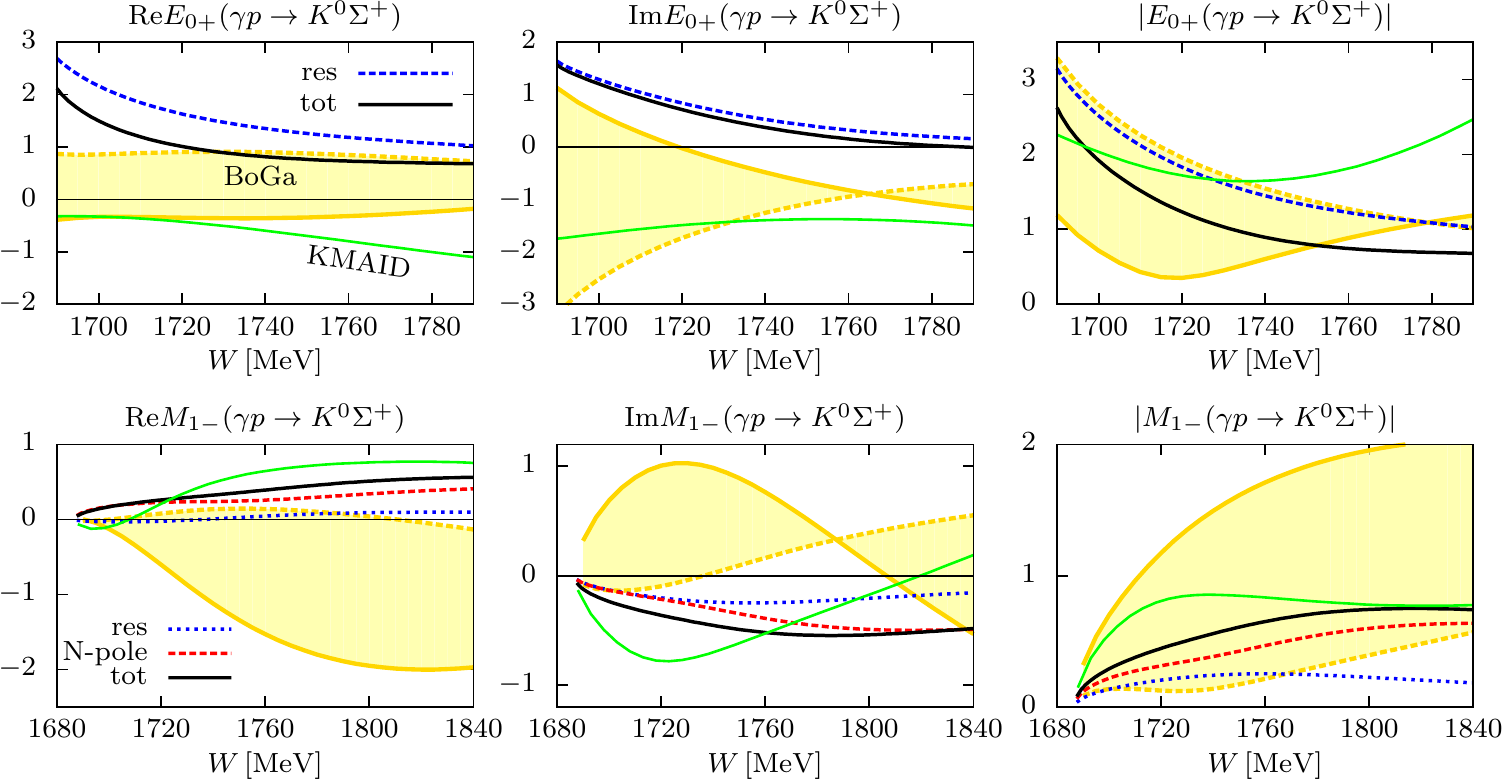}
\end{center}
\caption{(Color online.) The real and imaginary parts 
and the absolute values of the $E_{0+}$ and $M_{1-}$ 
photoproduction multipoles in the $K^0\Sigma^+$ channel.  
Notation as in fig.~\ref{fig:photoeta2}.}
\label{fig:photoKSa}
\end{figure*}
\vspace{-24pt}
\begin{figure*}[h!]
\begin{center}
\includegraphics[width=155mm]{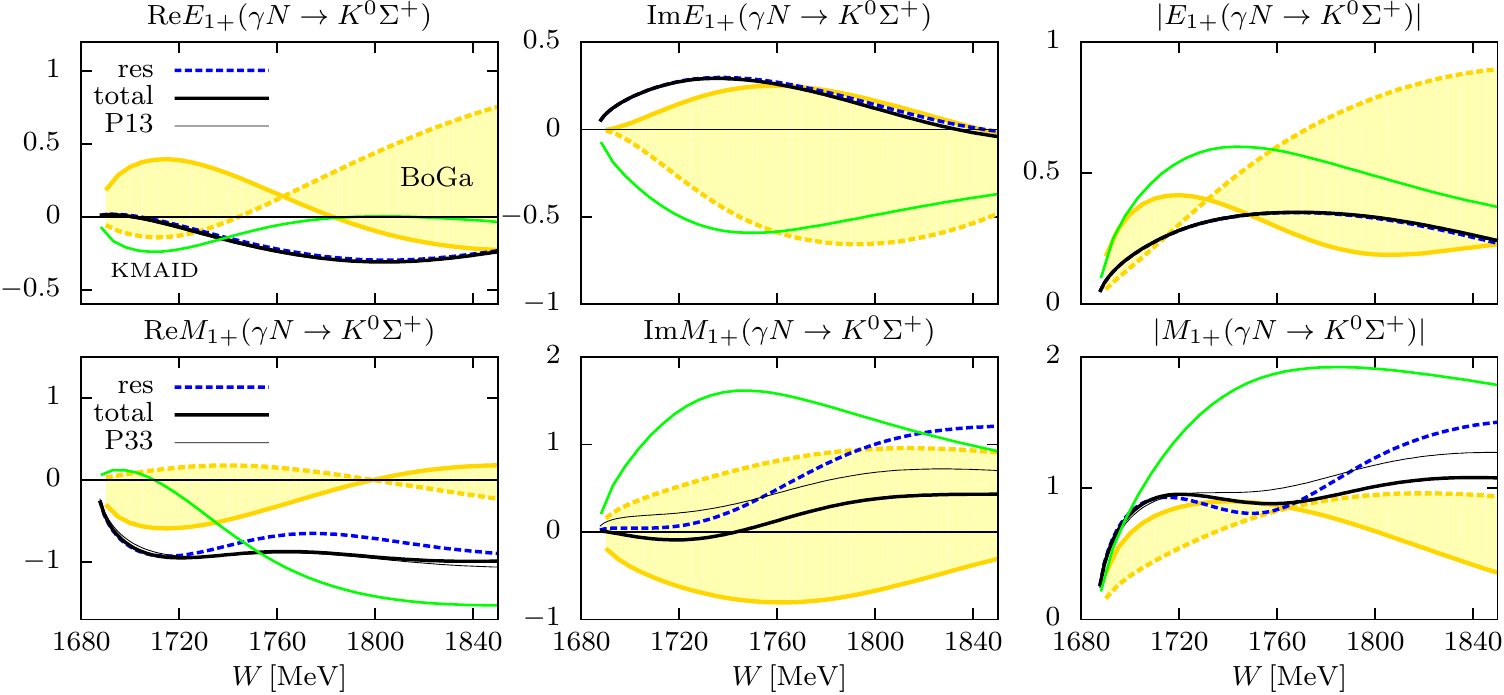}
\end{center}
\caption{(Color online.) The real and imaginary parts 
and the absolute values of the $E_{1+}$ and $M_{1+}$ photoproduction 
multipoles in the $K^0\Sigma^+$ channel.  
Notation as in fig.~\ref{fig:photoeta2}.
For the $E_{1+}$  ($M_{1+}$) multipole the leading $P_{13}$ ($P_{33}$)
contribution is also shown (thin lines).}
\label{fig:photoKSb}
\end{figure*}

\clearpage

\twocolumn

In fig.~\ref{fig:photoKSa} we display our results only for
the isospin-$1/2$ contribution and compare them to the relatively 
old Kaon MAID analysis~\cite{kaonMAID} and to the recent, 
but still inconclusive, analysis of the Bonn-Gatchina group.
The situation for the $M_{1-}$ multipole is similar;
the contribution of the $P_{31}$ wave is negligible while
the nucleon pole provides the dominant contribution.
Let us notice that in our calculation only the 
$\Delta(1910)\half^+$ is assumed in this partial wave.
In the case of the $M_{1+}$ and $E_{1-}$ amplitudes, 
both $P_{13}$ and $P_{33}$ partial waves contribute, however,
in magnetic dipole transition the $P_{33}$ wave strongly 
dominates in analogy with pion photoproduction 
in the region of the $\Delta(1232)$ resonance.
The situation is reversed for the $E_{1+}$ multipole; 
it is dominant in the $P_{13}$ wave since it involves 
an electric quadrupole transition of an $s$-wave quark to 
the $d$-state.

The role of the background terms in $K\Sigma$ is more important 
than in $K\Lambda$ production.
They are dominated by the coupled-channel effects in which 
the amplitudes for the $t$-channel $\gamma p\to\pi N$ and 
$\gamma p\to\pi\Delta$ processes couple to the corresponding 
$T_{\pi N\,K\Sigma}$ and $T_{\pi\Delta\,K\Sigma}$ matrix elements.
The background terms have the opposite sign with respect
to the resonant ones and weaken most of the amplitudes
in the $K^0\Sigma^+$ channel with respect to those in the  
$K^+\Sigma^0$ channel, in agreement with the Giessen group
analysis~\cite{Shklyar13}.

\begin{figure}[h!]
\begin{center}
\includegraphics[width=70mm]{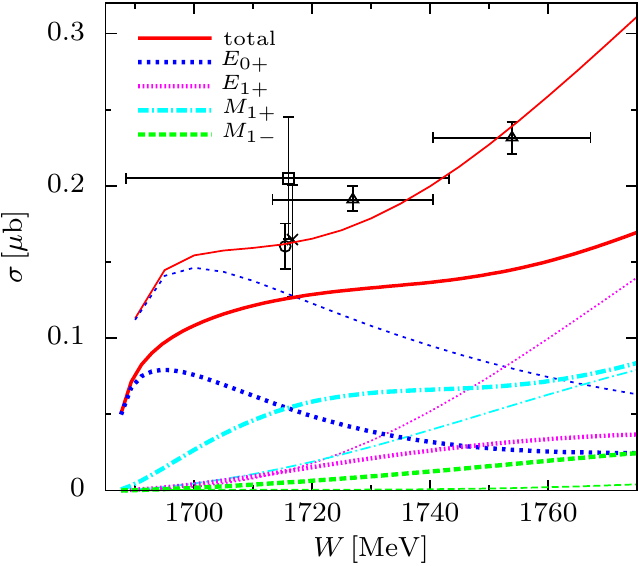}
\end{center}
\caption{(Color online.) Total cross-section for 
$\gamma p\to K^0\Sigma^+$. 
Experimental data are from~\cite{lawall05,cast08,ewald12,aguar13}.
Notation as in fig.~\ref{fig:sigmaKL}.}
\label{fig:sigmaKS}
\end{figure}

Different contributions to the  total cross-section are
displayed in fig.~\ref{fig:sigmaKS}.
Our results agree with the experimental values in the region
close to the threshold and unambiguously favour the  
Bonn-Gatchina 2014-1 solution over the 2014-2 one.
Not surprisingly, the $s$-wave kaons clearly dominate the 
threshold behaviour also in this channel.
At higher energies, the $P_{33}$ and the $P_{13}$ waves become 
the leading terms followed by the $P_{11}$ wave, in
agreement with the Bonn-Gatchina as well as the Giessen
\cite{Shklyar13} predictions.

\section{Conclusions}

We have proposed a relatively simple explanation for 
the resonance-like structure in the $\gamma n\to\eta n$ process 
at 1685~MeV as a consequence of the property of the relativistic
quark model which distinguishes between orbits with different $j$
on the one hand, and on the other hand as a coupled-channel effect 
due to the opening of the $K\Sigma$ channel.
Note that the former effect could not be obtained in a
non-relativistic constituent-quark model.

We have obtained a good overall agreement with observables
involving $\eta N$, $K\Lambda$ and $K\Sigma$ channels by using
a very small number of free parameters, which permits
us to conclude that the low-lying resonances in the considered
partial waves indeed fit very well in the quark-model picture.
Furthermore, most excitations and decays can be well described
in terms of single-quark transitions.

We have confirmed the leading role of the two lowest $S_{11}$ 
resonances in the threshold region found in several partial-wave
analyses, as well as the relevance of the $P_{11}$, $P_{13}$ waves 
at somewhat higher energies for the isospin-$1/2$ processes, and 
the $P_{33}$ wave for isospin-$3/2$ processes, while 
the $S_{31}$ and $D$-wave resonances play a minor role.

The SU(3) symmetry relations for the meson-baryon and photon-baryon 
coupling turn out to be well justified by the present calculation.

The resonance that may not fit in our scheme could be
the $N(1710)\half^+$.
In our model the dominant contribution in the $P_{11}$ partial
wave stems from the  $N(1440)\half^+$ and the nucleon pole.
If it turns out that the $N(1710)\half^+$ resonance indeed 
dominates the production of $\eta$ mesons and kaons, our 
assumption about the single-quark transition mechanism does 
not apply and one should seek an explanation beyond the simple 
quark model for the structure of this resonance.


\appendix
\section{\label{P13} The $P_{13}$ resonances}

The $LS$-coupled states belonging to the ${\bf 70}$ multiplet
recoupled to the $jj$-coupled basis read:
\begin{eqnarray*}
  |{\bf 70}, {}^4{\bf 8}, J=\thalf\rangle
  &=& -{2\over\sqrt5} \;|{\bf{8}},\thalf,SSA\rangle
      + {1\over\sqrt5}\;|{\bf{8}},\thalf,SSB\rangle \>,
\\
  |{\bf 70}, {}^2{\bf 8}, J=\thalf\rangle
  &=& -{1\over\sqrt{10}} \;|{\bf{8}},\thalf,SSA \rangle
      -{1\over\sqrt2}    \;|{\bf{8}},\thalf,SSA \rangle'
\\ &&
      -\sqrt{2\over5}    \;|{\bf{8}},\thalf,SSB \rangle \>,
\end{eqnarray*}
where $S$ stands for the $s$-quark orbit and $A$ and $B$ for
$j=\thalf$ and $j=\fhalf$ orbits, respectively.
The state with total $J=\thalf$ and $M_J=\thalf$ 
is constructed as 
\begin{eqnarray*}
|{\bf 8},\thalf, SSX\rangle &=& {1\over\sqrt2}
 \left[\phi^{[21]}_{11}|\thalf\thalf,SSX\rangle^{[21]}_{11}
\right.\\ && \left.
    + \phi^{[21]}_{21}|\thalf\thalf,SSX\rangle^{[21]}_{21}
\right] \>,
\end{eqnarray*}
where 
\begin{eqnarray*}
|\thalf\thalf, SSA\rangle^{[21]}_{11} &=& {1\over\sqrt{60}}
\bigl[2(2\U\U\beta - \U\beta\U - \beta\U\U)\phantom{\sqrt3}\bigr. 
\\ && 
   -\sqrt3(2\U\D\alpha+2\D\U\alpha
\\ && \bigl.
   -\U\alpha\D-\alpha\U\D-\D\alpha\U-\alpha\D\U)\bigr] \>,
\\
|\thalf\thalf, SSA\rangle^{[21]}_{21} &=& {1\over\sqrt{20}}
\bigl[2(\U\beta\U - \beta\U\U)\bigr. 
\\ && \bigl.
  - \sqrt3(\U\alpha\D-\alpha\U\D+\D\alpha\U-\alpha\D\U)\bigr] \>,
\\
|\thalf\thalf, SSA\rangle^{[21]'}_{12} &=& {1\over2}
\bigl[\U\alpha\D+\alpha\U\D-\D\alpha\U-\alpha\D\U\bigr] \>,
\\
|\thalf\thalf, SSA\rangle^{[21]'}_{22} &=& {1\over\sqrt{12}}
\bigl[2\U\D\alpha-2\D\U\alpha
\bigr. \\ && \bigl.
 +\U\alpha\D-\alpha\U\D+\alpha\D\U-\D\alpha\U\bigr] \>,
\\
|\thalf\thalf, SSB\rangle^{[21]}_{11} &=& {1\over\sqrt{90}}
\bigl[\sqrt{10}(2\D\D a - \D a\D - a\D\D)\bigr.
\\ && 
        -\sqrt2(2\U\D b+2\D\U b
\\ &&
       -\U b\D-b\U\D-\D b\U-b\D\U)
\\ && \bigl.
   + (2\U\U c - \U c\U - c\U\U)\bigr] \>,
\\
|\thalf\thalf, SSB\rangle^{[21]}_{21} &=& {1\over\sqrt{30}}
\bigl[\sqrt{10}(\D a\D - a\D\D)\bigr.
\\ && 
   -\sqrt2(\U b\D+b\U\D-\D b\U-b\D\U)
\\ && \bigl.
   + (\U c\U - c\U\U)\bigr] \>,
\end{eqnarray*}
and
$\U=(j=\half,m_j=\half)$, $\D=(\half,-\half)$;
\\
$\alpha =(j=\thalf,m_j=\thalf)$,
$\beta =(\thalf,\half)$;
\\
$a=(j=\fhalf,m_j=\fhalf)$,
$b =(\fhalf,\thalf)$, $c =(\fhalf,\half)$\,.\\
Here $\phi^{[21]}_{st}$ are the flavour wave functions:
$\phi^{[21]}_{11}=\break\sqrt{1/6}\left(2uud-udu-duu\right)$,
$\phi^{[21]}_{21}=\sqrt{1/2}\left(udu-duu\right)$.

\section{The meson-quark vertices}

The meson-quark vertices $V^M(k)$ in (\ref{defK}) are evaluated in the 
Cloudy Bag Model assuming that one of the three quarks is excited 
from the $1s$ orbit to an $l$-wave orbit with $j=l\pm \half$. 

For the pion with orbital momentum $L$, its third component $M$
and the third component of isospin $t$ we have
\begin{eqnarray*}
V^\pi_{LMt}(k) 
&=&
{\sqrt{3}\over 2\pi f}\,\sum_{l,j=l\pm\half} \sqrt{2l+1}\,
  \CG{1}{0}{l}{0}{L}{0} W(Ll\half\half;1j)
\nonumber\\ 
&\times&
  (-1)^{l+\h-j}
  \sqrt{\omega_s\omega_{lj}\over(\omega_s-1)(\omega_{lj}\mp(j+\half))}
\nonumber\\ 
&\times&
{k^2\over\sqrt{2\omega_k}}\,{j_L(kR_\mathrm{bag})\over kR_\mathrm{bag}}\,
   \sum_{i=1}^3\Sigma^{[j\h]}_{LM}(i)\tau_t(i)\,,
\end{eqnarray*}
where
$$
\langle jm_j|\Sigma^{[j\h]}_{LM}|\half m_s\rangle
=\CG{\h}{m_s}{L}{M}{j}{m_j}\,,
\qquad
 \Sigma^{[\h\h]}_{1M} = {1\over\sqrt3}\;\sigma_M\,,
$$
and $\omega_{lj}$ is a solution of the boundary condition 
equation for massless quarks: 
$j_l(\omega_{lj}R_\mathrm{bag})=\pm j_{l\pm1}(\omega_{lj}R_\mathrm{bag}),
j=l\pm\half$; where $j_l$ are the spherical Bessel functions, 
and $W(\ldots)$ are the Racah coefficients.

For the $\eta$ meson the $\tau_t(i)$ operator is replaced by
the $\lambda_8(i)$ operator, and for the kaons by
the $V_t(i)+U_t(i)$ operator.

In our approach the state of the $\rho$ meson is specified by 
its total angular momentum $J_\rho$ and its third component $M$, 
orbital angular momentum $L$, and isospin $t$.
It is related to the state in which the total spin $S$ of the
$\rho N$ system is specified:
$$
 |SLJM_J\rangle\kern-3pt =\kern-4pt
\sum_{J_\rho}\kern-4pt\sqrt{2S+1}\sqrt{2J_\rho+1}\,
  W(LJ1\half;SJ_\rho) |J_\rho LJM_J\rangle,
$$
where $J$ and $M_J$ refer to the total angular momentum of
the $\rho N$ system.
The corresponding vertex operator is given by
\begin{eqnarray*}
 V^{\rho}_{J_\rho\,L M t}(k)
&=&
   {3\over2\pi f_\rho}\,(-1)^{J_\rho+L}\kern-6pt
  \sum_{l,j=l\pm\half}\kern-6pt\sqrt{2J_\rho+1}\sqrt{2l+1}\,\CG{1}{0}{l}{0}{L}{0}
\nonumber\\ 
&\times&
W(lL11;1J_\rho) W(J_\rho l \half\half;1j)
{k^2\over\sqrt{2\omega_k}}\,{j_{L}(kR_\mathrm{bag})\over kR_\mathrm{bag}}
\nonumber\\
&\times&
\sqrt{\omega_s\omega_{lj}\over(\omega_s-1)(\omega_{lj}\mp(j+\half))}
\sum_{i=1}^3\Sigma^{[j\h]}_{J_\rho M}(i)\tau_t(i)\,.
\end{eqnarray*}

\section{\label{EMmulti} The E and M multipole operators}

We give here the expressions  for 
the quark part of the transverse magnetic and electric multipole 
operators involving transition from an $s$-wave quark orbit
to an $l$-wave quark orbit with $j=l\pm\half$; 
$u_s$, $v_s$ and $u_{lj}$, $v_{lj}$ are the corresponding bispinors.
%
%
\begin{eqnarray}
\int\d\vec{r}\,\vec{j}^q\cdot\vec{A}_{LM}^m
&=&  \i^{L+1}\,
     \sqrt{2L+1}\int\d r\,r^2j_L(qr)
\nonumber\\ &&\kern-56pt\times     \sum_{l,j=l\pm\half}
\biggl\{(-1)^{L+j-\h}
      \CG{1}{0}{L}{0}{L}{0}\;\delta_{l,L}\,(u_{lj}v_s-v_{lj}u_s)\biggr.
\nonumber\\
&&\kern-56pt 
+6\sqrt{2L+1}\sqrt{2l+1}\,W(lL11;1L)W(Ll\h\h;1j)
\nonumber\\
&&\kern-56pt\biggl.
\times\CG{1}{0}{l}{0}{L}{0}(u_{lj}v_s+v_{lj}u_s)\biggr\}
\left[{1\over6} + {1\over2}\,\tau_0(i)\right]\Sigma_{LM}^{[j\h]} \>,
\nonumber\\[24pt]
%
%
\int\d\vec{r}\,\vec{j}^q\cdot\vec{A}_{L1}^e
&=&  \i^{L}\,
\int\d r\,r^2\sum_{l,j=l\pm\half}\biggl\{\sqrt{L+1}\,j_{L-1}(qr)\biggr.
\nonumber\\
&&\kern-60pt 
  \times \biggl[
  (-1)^{L+j+\h}\CG{1}{0}{L}{0}{L-1}{0}\;
  \delta_{l,L}\,(u_{lj}v_s-v_{lj}u_s)\biggr.
\nonumber\\
&& \kern-56pt 
  - 6\sqrt{2L+1}\sqrt{2l+1}\,W(lL-1\,11;1L)W(Ll\h\h;1j)
\nonumber\\
&&\biggl.\kern-56pt
\times\CG{1}{0}{l}{0}{L-1}{0}(u_{lj}v_s+v_{lj}u_s)\biggr]
\nonumber\\
&&\kern-56pt 
  - \sqrt{L}\,j_{L+1}(qr)\biggl[
  (-1)^{L+j+\h}\CG{1}{0}{L}{0}{L+1}{0}\;
  \delta_{l,L}\,(u_{lj}v_s-v_{lj}u_s)\biggr.
\nonumber\\
&&\kern-56pt 
 - 6\sqrt{2L+1}\sqrt{2l+1}\,W(lL+1\,11;1L)W(Ll\h\h;1j)
\nonumber\\
&&\kern-56pt\biggl.\biggl.
\times\CG{1}{0}{l}{0}{L+1}{0}(u_{lj}v_s+v_{lj}u_s)\biggr]\biggr\}
\left[{1\over6} + {1\over2}\,\tau_0(i)\right]\Sigma_{L1}^{[j\h]}\,,
\nonumber
\end{eqnarray}
where $q=|\vec{q}|$ is the photon momentum.

For the photon-pion interaction the corresponding
operators are written in the form
%
%
\begin{eqnarray*}
\int\d\vec{r}\,\vec{j}_\pi\cdot\vec{A}_{LM}^m
&=&
-\i^{L+1}\sqrt2\sum_{ll'}(-1)^{L-1+l-l'}
\sqrt{2l+1} \;
\\ &&\kern-66pt \times
{2\over\pi}\int\d r\,r^2
\int{\d k\,  k    \over\sqrt{2\omega_k}}
\int{\d k'\,{k'}^2\over\sqrt{2\omega_k'}} j_l(kr) j_L(qr)
\\ &&\kern-66pt \times 
\left[j_{l'\kern-1pt+\kern-1pt1}(k'r)
\sqrt{(l'\kern-1pt+\kern-1pt1)(2l'\kern-1pt+\kern-1pt3)}
\CG{l}{0}{l'\kern-1pt+\kern-1pt1}{0}{L}{0}
W(l'1lL;l'\kern-1pt+\kern-1pt1,L)\right.
\\
   && \kern-62pt \left. + j_{l'-1}(k'r)\sqrt{l'(2l'-1)}
   \CG{l}{0}{l'-1}{0}{L}{0}W(l'1lL;l'-1,L)\right]
\\
   && \kern -66pt\times   {\mathbf{A}_{ll'}}^{10}_{LM}\,,
\end{eqnarray*}
%
%
\begin{eqnarray*}
\int\d\vec{r}\,\vec{j}_\pi\cdot\vec{A}_{L1}^e
&=&
-\i^{L+2}\sum_{ll'}(-1)^{L-1+l-l'}{\sqrt2\sqrt{2l+1}\over\sqrt{L(L+1)}} 
\\ &&\kern-66pt \times
{2\over\pi}\int\d r\,r^2
\int{\d k\,  k    \over\sqrt{2\omega_k}}
\int{\d k'\,{k'}^2\over\sqrt{2\omega_k'}} j_l(kr)
\\
&& \kern-66pt\times  \left\{ -L\sqrt{L+1}\,  j_{L+1}(qr) \right.
\\
 && \kern-30pt\times 
\left[j_{l'+1}(k'r)\sqrt{(l'+1)(2l'+3)}\CG{l}{0}{l'+1}{0}{L+1}{0}\right.
 \\ && \times       W(l'1lL+1;l'+1,L)
\\
   && \kern-26pt + j_{l'-1}(k'r)\sqrt{l'(2l'-1)}\CG{l}{0}{l'-1}{0}{L+1}{0}
\\ && \times \Bigl. W(l'1lL+1;l'-1,L)\Bigr]
\\
&& \kern-66pt    +(L+1)\sqrt{L}\,  j_{L-1}(qr)
\\
&&    \kern-30pt\times 
  \left[j_{l'+1}(k'r)\sqrt{(l'+1)(2l'+3)}\CG{l}{0}{l'+1}{0}{L-1}{0}\right.
\\ && \times        W(l'1lL-1;l'+1,L)
\\ 
   && \kern-26pt
        + j_{l'-1}(k'r)\sqrt{l'(2l'-1)}\CG{l}{0}{l'-1}{0}{L-1}{0}
\\ && \times \Bigr.\Biggr.  W(l'1lL-1;l'-1,L)\Bigr]\Bigr\}
\\
   && \kern -66pt\times   {\mathbf{A}_{ll'}}^{10}_{L1}
\end{eqnarray*}
Here $ {\mathbf{A}_{ll'}}^{L1}_{10}$ is the product of the $l$- 
and $l'$-wave pion creation and annihilation operators coupled 
to angular momentum $L$ and isospin 1.
Note that the orbital angular momentum of the pion can be changed.
For the kaons, the expressions differ only in the coefficients
of the isospin coupling.


\begin{thebibliography}{99} 

\bibitem{EPJ2005} 
P. Alberto, L. Amoreira, M. Fiolhais, B. Golli, 
and S. \v{S}irca, Eur. Phys. J. A {\bf 26}, 99 (2005).
\bibitem{EPJ2008} 
B. Golli  and S. \v{S}irca, 
  Eur. Phys. J. A {\bf 38}, 271 (2008).
\bibitem{EPJ2009} 
B. Golli, S. \v{S}irca, and M. Fiolhais,
  Eur. Phys. J. A {\bf 42}, 185 (2009).
\bibitem{EPJ2011} 
B. Golli, S. \v{S}irca,
  Eur. Phys. J. A {\bf 47}, 61 (2011).
\bibitem{EPJ2013} 
B. Golli, S. \v{S}irca,
  Eur. Phys. J. A {\bf 49}, 111 (2013).


\bibitem{crede05} V.~Cred\'e {\em et al.} (CBELSA Collaboration),
  Phys. Rev. Lett. {\bf 94}, 012004 (2005).
\bibitem{bartalini07} O.~Bartalini {\em et al.} (GRAAL Collaboration),
  Eur. Phys. J. A {\bf 33}, 169 (2007).
\bibitem{crede09} V.~Cred\'e {\em et al.} (CBELSA/TAPS Collaboration),
  Phys. Rev. C {\bf 80}, 055202 (2009).
\bibitem{williams09} M.~Williams {\em et al.} (CLAS Collaboration),
  Phys. Rev. C {\bf 80}, 045213 (2009).
\bibitem{mcnicoll10} E.~F.~McNicoll {\em et al.} (Crystal Ball Collaboration),
  Phys. Rev. C {\bf 82}, 035208 (2010).
\bibitem{werth14} D.~Werthm\"uller {\em et al.} (A2 Collaboration), 
  Phys. Rev. C {\bf 90}, 015205 (2014).


\bibitem{rebre05} D.~Rebreyend {\em et al.} (GRAAL Collaboration),
  Int. J. Mod. Phys. A {\bf 20}, 1554 (2005).
\bibitem{jaegle08} I.~Jaegle {\em et al.} (CBELSA/TAPS Collaboration),
  Phys. Rev. Lett. {\bf 100}, 252002 (2008).
\bibitem{jaegle11} I.~Jaegle {\em et al.} (CBELSA/TAPS Collaboration),
  Eur. Phys. J. A {\bf 47}, 89 (2011).

\bibitem{Kuznetsov07} V. Kuznetsov {\em et al.}, 
Phys. Lett. B {\bf 647}, 23 (2007).
\bibitem{Azimov05}
Ya. Azimov {\em et al.} Eur. Phys. J. A {\bf 25},  325 (2005).
\bibitem{Doering10}
 M.~D\"oring, K. Nakayama, Phys. Lett. B {\bf 683}, 145 (2010).
\bibitem{Shyam08} 
R. Shyam and O. Scholten, Phys. Rev. C {\bf 78}, 065201 (2008).
\bibitem{Shklyar07}
V. Shklyar, H. Lenske, U. Mosel, Phys. Lett. B {\bf 650}, 172 (2007).
\bibitem{Shklyar13}
V. Shklyar, H. Lenske, U. Mosel, Phys. Rev. C {\bf 87}, 015201 (2013).
\bibitem{BoGa09}
A.~V.~Anisovich {\em et al.},  Eur. Phys. J. A {\bf 41},  13 (2009).
\bibitem{BoGa15}
A.~V.~Anisovich {\em et al.},  Eur. Phys. J. A {\bf 51},  72 (2015).
\bibitem{Zhong11}
X.-H. Zhong, Q. Zhao, Phys. Rev. C {\bf 84}, 045207 (2011).
\bibitem{Saghai09} 
Jun He, B. Saghai, Phys. Rev. C {\bf 80}, 015207 (2009).
\bibitem{Saghai10} 
Jun He, B. Saghai, Phys. Rev. C {\bf 82}, 035206 (2010).
\bibitem{MaiPLB11} D. Rui\'c, M. Mai, U.-G. Mei\ss ner,
Phys. Lett. B {\bf 704} 659 (2011).
\bibitem{MaiPRD12} M. Mai, P. C. Bruns, and U.-G. Mei\ss ner,
Phys. Rev. D {\bf 86}, 094033 (2012).


\bibitem{bock94} M.~Bockhorst {\em et al.} (SAPHIR Collaboration), 
  Z. Phys. C {\bf 63}, 37 (1994).
\bibitem{tran98} M.~Q.~Tran {\em et al.} (SAPHIR Collaboration), 
  Phys. Lett. B {\bf 445}, 20 (1998).
\bibitem{glan04} K.~H.~Glander {\em et al.} (SAPHIR Collaboration), 
  Eur. Phys. J. A {\bf 19}, 251 (2004).
\bibitem{brad04} R.~Bradford {\em et al.} (CLAS Collaboration), 
  Phys. Rev. C {\bf 73}, 035202 (2006).
\bibitem{nabb04} J.~W.~C. McNabb {\em et al.} (CLAS Collaboration), 
  Phys. Rev. C {\bf 69},  042201(R) (2004).
\bibitem{sumi06} M.~Sumihama {\em et al.} (LEPS Collaboration), 
  Phys. Rev. C {\bf 73}, 035214 (2006).
\bibitem{ller07} A.~Lleres {\em et al.} (GRAAL Collaboration), 
  Eur. Phys. J. A {\bf 31}, 79 (2007).
\bibitem{ller09} A.~Lleres {\em et al.} (GRAAL Collaboration), 
  Eur. Phys. J. A {\bf 39}, 149 (2009).
\bibitem{crack10} M.~E.~McCracken {\em et al.} (CLAS Collaboration), 
  Phys. Rev. C {\bf 81}, 025201 (2010).


\bibitem{Li95} 
Z. Li, Rev. C {\bf 52}, 1648 (1995).
\bibitem{Li96}
Z. Li, M. Wei-Hsing, Z. Lin, Rev. C {\bf 54}, R2171 (1996).
\bibitem{Ulf97} 
S. Steininger and Ulf-G. Mei{\ss}ner, 
Phys. Lett. B {\bf 391}, 446 (1997).
\bibitem{Ulf06}
B. Borasoy, P. C. Bruns, U.-G. Meissner, and R. Nissler, 
Eur. Phys. J. A {\bf 34}, 161 (2007).
\bibitem{Shklyar05}
V. Shklyar, H. Lenske, U. Mosel, Phys. Rev. C {\bf 72}, 015210 (2005).
\bibitem{Shyam10} 
R. Shyam, O. Scholten, and H. Lenske, Phys. Rev. C {\bf 81}, 015201 (2010).
\bibitem{Gent2012}
L. De Cruz, J. Ryckebusch, T. Vrancx, and P. Vancraeyveld,
 Phys. Rev. C {\bf 86}, 015212 (2012).
\bibitem{Mart10} 
T. Mart, Phys. Rev. C {\bf 82}, 025209 (2010).
\bibitem{Mart11} 
T. Mart, Phys. Rev. C {\bf 83}, 048203 (2011).
\bibitem{Manley12b}
M. Shrestha and D. M. Manley,  Phys. Rev. C {\bf 86}, 045204 (2012).
\bibitem{Manley12a}
M. Shrestha and D. M. Manley,  Phys. Rev. C {\bf 86}, 055203 (2012).




\bibitem{goers99} S.~Goers {\em et al.} (SAPHIR Collaboration),
  Phys. Lett. B {\bf 464}, 331 (1999).
\bibitem{lawall05} R.~Lawall {\em et al.} (SAPHIR Collaboration), 
  Eur. Phys. J. A {\bf 24}, 275  (2005).
\bibitem{klein05} F.~J.~Klein {\em et al.} (CLAS Collaboration),
  Nucl. Phys. A {\bf 754},  321(2005).
\bibitem{cast08} R.~Castelijns {\em et al.} (CBELSA/TAPS Collaboration), 
  Eur. Phys. J. A {\bf 35},  39 (2008).
\bibitem{ewald12} R.~Ewald {\em et al.} (CBELSA/TAPS Collaboration), 
  Phys. Lett. B {\bf 713}, 180 (2012).
\bibitem{aguar13} P.~Aguar-Bartolom\'e {\em et al.} (A2 Collaboration), 
  Phys. Rev. C {\bf 88}, 044601 (2013).

\bibitem{BoGa13ks} 
A.~V.~Anisovich {\em et al.},  Eur. Phys. J. A {\bf 49},  158 (2013).
\bibitem{Mart14} 
T. Mart, Phys. Rev. C {\bf 90}, 065202 (2014).
\bibitem{Roenchen13} 
D. R\"onchen {\em et al.},  Eur. Phys. J. A {\bf 49},  44 (2013).
\bibitem{Kamano13} 
H.~Kamano, S.~X.~Nakamura, T.-S.~H.~Lee, T.~Sato,
Phys. Rev. C {\bf 88}, 035209 (2013).
\bibitem{Thomas85} 
E. A. Veit, B. K. Jennings, A. W. Thomas, R. C. Barret,  
Phys. Rev. D {\bf 31}, 1033 (1985).
\bibitem{gaugeCBM} Gerald A. Miller, and Anthony W. Thomas,
Phys. Rev. C {\bf 56}, 2329 (1997).
\bibitem{CBM4Delta} D. H. Lu, A. W. Thomas, and A. G. Williams,
Phys. Rev. C {\bf 55}, 3108 (1997).
\bibitem{Isgur77} 
N. Isgur and G. Karl. Phys. Lett. {\bf 72B}, 109 (1977).
\bibitem{deGrand76b} 
T. A. deGrand, Ann. Phys. {\bf 101}, 496 (1976).
\bibitem{Myhrer84b} 
F. Myhrer and J. Wroldsen, Z. Phys. C 25, 281 (1984).
\bibitem{Moorhouse1966}
R. G. Moorhouse, Phys. Rev. Lett. {\bf 16} 772 (1996).
\bibitem{Arndt06} 
R. A. Arndt, W. J. Briscoe, I. I. Strakovsky, 
 and R. L. Workman, Phys. Rev. C {\bf 74}, 045205 (2006).
\bibitem{SAID} 
\verb|http://gwdac.phys.gwu.edu/analysis/pr_analysis.html|.
\bibitem{Pedro} 
P. Alberto, M. Fiolhais, B. Golli, and J. Marques,
Phys. Lett. B {\bf 523}, 273 (2001).
\bibitem{BoGa-data}
  \verb|http://pwa.hiskp.uni-bonn.de/|.
\bibitem{PLB96} 
M.~Fiolhais, B.~Golli, S.~\v{S}irca, Phys. Lett. B {\bf 373}, 229 (1996).
\bibitem{Manley92} 
D.~M.~Manley, E. M.~Saleski,  Phys. Rev. D {\bf 45}, 4002 (1992).
\bibitem{ltpc} 
L.~Tiator, $\eta$MAID 2015, private communication.
\bibitem{kaonMAID} 
F.~X.~Lee, T.~Mart, C.~Bennhold, H.~Haberzettl, 
  L.~E.~Wright, Nucl. Phys. A {\bf 695}, 237 (2001);
  see also \verb|http://portal.kph.uni-mainz.de/MAID/kaon/|.

\end{thebibliography}
\end{document}